\pdfoutput=1
\documentclass{article}

\usepackage{arxiv}
\usepackage[utf8]{inputenc}
\usepackage[T1]{fontenc}
\usepackage{hyperref}
\usepackage{url}
\usepackage{booktabs}
\usepackage{amsfonts}
\usepackage{nicefrac}
\usepackage{microtype}
\usepackage{lipsum}
\usepackage{graphicx}
\usepackage[numbers]{natbib}
\usepackage{doi}
\usepackage{hyperref}
\usepackage{url}
\usepackage{subcaption}
\usepackage{amsmath} 
\usepackage{multirow}
\usepackage{algorithm}
\usepackage{algorithmic}
\usepackage{listings}
\usepackage{footnote}
\usepackage{booktabs}
\usepackage{makecell}
\usepackage{multirow}
\usepackage{tabu}
\usepackage{threeparttable}
\usepackage{tabularx}
\usepackage{tabulary}
\usepackage{graphicx}
\usepackage{xspace}
\usepackage{adjustbox}
\usepackage{cleveref}
\usepackage{enumitem}

\newif\ifcomments
\commentstrue
\ifcomments
    
\else
    \providecommand{\ion}[1]{}
\fi

\long\def\eat#1{}

\title{Specifications: The missing link to making the development of LLM systems an engineering discipline}

\author{%
  Ion Stoica$^{1}$, Matei Zaharia$^{1}$, Joseph Gonzalez$^{1}$, Ken Goldberg$^{1}$, Koushik Sen$^{1}$, Hao Zhang$^{2}$,\\[4pt]
  \textbf{Anastasios N. Angelopoulos$^{1}$, Shishir G. Patil$^{1}$, Lingjiao Chen$^{3,4}$, Wei-Lin Chiang$^{1}$, Jared Q. Davis$^{3}$} \\[4pt]
  $^{1}$UC Berkeley, $^{2}$UC San Diego, $^{3}$Stanford University, $^{4}$Microsoft Research \\
}

\renewcommand{\headeright}{ }
\renewcommand{\undertitle}{ }

\begin{document}
\maketitle

\begin{abstract}
Despite the significant strides made by generative AI in just a few short years, its future progress is constrained by the challenge of building \emph{modular} and \emph{robust} systems. This capability has been a cornerstone of past technological revolutions, which relied on combining components to create increasingly sophisticated and reliable systems. Cars, airplanes, computers, and software consist of components—such as engines, wheels, CPUs, and libraries—that can be assembled, debugged, and replaced.
A key tool for building such reliable and modular systems is \emph{specification}: the precise description of the expected behavior, inputs, and outputs of each component. However, the generality of LLMs and the inherent ambiguity of natural language make defining specifications for LLM-based components (e.g., agents) both a challenging and urgent problem. In this paper, we discuss the progress the field has made so far—through advances like structured outputs, process supervision, and test-time compute—and outline several future directions for research to enable the development of modular and reliable LLM-based systems through improved specifications.

\end{abstract}

\section{Introduction}
\label{sec:intro}

Software has been one of the main drivers of economic growth over the past several decades, a trend famously captured by Andreessen in his influential blog post, “Why software is eating the world,” more than a decade ago~\cite{why-software-is-eating-the-world}. More recently, AI—particularly Large Language Models (LLMs)—has emerged as the next revolution, poised to disrupt the very software that has been "eating the world"~\cite{nvidia-ceo-software-AI,AI-eating-software,AI-eating-software2}. However, to fully realize this vision, we need to build LLM-based systems with the same level of reliability and rigor as those found in established engineering disciplines, such as control theory, mechanical engineering, and software engineering.

\begin{figure}[h]
    \centering
    \includegraphics[width = 0.8\textwidth]{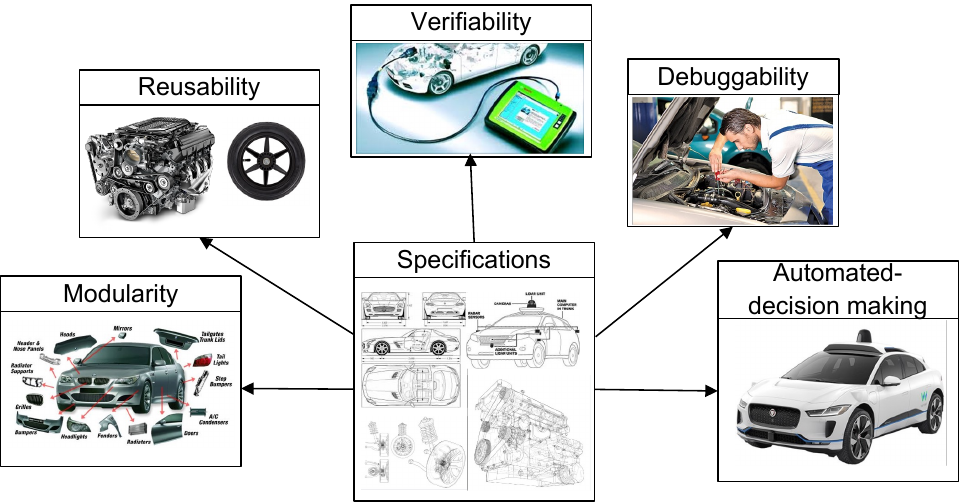}
    \caption{A car consists of many parts (modularity), where some parts are shared with other cars (reusability). When something goes wrong, the mechanic diagnoses the problem (verifiability) and then fixes it (debuggability). Furthermore, some functions of the car don’t require human intervention (automated decision making). To enable these properties we need detailed schematics and detailed descriptions of how each of these components works (specifications).
    }
    \label{fig:eng-properties-example} 
\vskip -0.1in
\end{figure}

One key tool that has historically enabled the rapid growth of engineering disciplines is \textit{specification}. Specifications describe the expected behavior, inputs, and outputs of a system. These specifications have different levels of precision and come in many forms, including formal specifications, product requirements documents (PRDs), and user manuals. Specifications help developers in several ways: (1) decomposing complex systems into smaller components, (2) reusing existing components when possible, (3) verifying that a system works as intended (i.e., it meets its specification), (4) fixing the system when it does not, and (5) creating systems capable of making decisions without human intervention. Figure~\ref{fig:eng-properties-example}  illustrates how specifications enable these five properties in the automotive industry.

In this paper, we distinguish between two types of specifications: (i) a description of what the task should do, which we call a \textbf{statement specification}, and (ii) a description that enables a user to verify task solutions (outputs), which we call a \textbf{solution specification}. Essentially, statement specifications answer the question, "What should the task accomplish?" while solution specifications answer the question, "How can one verify that the task’s solution adheres to the statement specification?"
In traditional software systems, the statement specification can be captured in a Product Requirements Document (PRD), whereas the solution specification can be expressed as input-output tests (e.g., unit tests). In formal frameworks like Coq/Gallina~\cite{coq,gallina}, the statement specification is represented by a formal specification, and the solution specification is captured by a proof demonstrating that the generated code is correct. When building a radio, the statement specification is represented by the circuit schematics, while the solution specification is defined by the expected voltage and amperage at different points in the circuit. 
Finally, in some cases, the statement and solution specification coincide, e.g., the task, "solve $x^3 - x = 0$", specifies both what the task should do (i.e., solve the equation) and how a solution should be verified—by substituting the solution back into the equation to check the equality.

Defining good specifications is challenging. Apart from formal specifications, most specifications contain some degree of ambiguity. For example, it is notoriously difficult to describe and test for every possible behavior of a software system~\cite{brooks1975mythical}. That said, LLMs are even more ambiguous than software systems, and this has significant implications. With LLMs, users provide specifications through natural language prompts. While natural language is highly general and flexible, it often encourages users to specify ambiguous, ill-defined tasks.\footnote{This should come as no surprise, as natural language was developed by humans to communicate with one another. Humans excel at handling ambiguity by relying on rich context to complement language, including facial expressions, tone of voice, and shared historical and cultural backgrounds. Additionally, humans often ask clarifying questions to resolve ambiguity.} While encouraging ambiguity has been one of the main strengths of LLMs, this has also led to spectacular failures, such as the well-known "hallucinations", where LLMs generate incorrect answers to user input prompts~\cite{hallucinations, code-hallu, healthcare-hallucination}.

As illustrated by "hallucinations," the lack of clear specifications makes it challenging to build \emph{reliable systems}. If a task is not well-specified, how can one even determine whether the task is being performed correctly? Moreover, this ambiguity complicates the development of \emph{modular systems} composed of multiple LLM components (e.g., agents). Without a clear understanding of what each component should do, how can one define the behavior of a system that integrates these components? As we will argue in Section~\ref{sec:challenges}, the challenge of building reliable, modular systems is significantly hindering the growth of the LLM ecosystem. Finally, in the absence of well-defined solution specifications, LLM-based applications (e.g., Q\&A, customer support, copilot tools) often depend on humans to evaluate the quality of their outputs, as essential context is frequently missing from the task specifications. This dependence makes such applications unsuitable for fully automated decision-making without human intervention.

\begin{table}[h]
\begin{tabular}{|p{2cm}|p{7cm}|p{6.5cm}|}
\hline
\textbf{Failures} &
  \textbf{Description \& Impact} &
  \textbf{Possible fixes}\\ \hline
  Wrong policy &
  An Air Canada customer service chatbot provided wrong policy information about a ticket being eligible for refund. Eventually, Air Canada was ordered to honor the erroneous refund policy ~\cite{air-canada-hallucination}. & 
Specify policies to verify the answers. Ask customers for clarifications to disambiguate their questions, if needed. \\ \hline
  Wrong offer & A Chevrolet customer service chatbot agreed to sell to a customer a late model for \$1~\cite{chevrolet-1-dollar}. & 
  Specify the rules for a valid offer (e.g., price $\geq$ threshold), and monitor generated offers. \\ \hline
  Wrong health advice & 
  LLMs provided incorrect and harmful health advice~\cite{healthcare-hallucination}. & Specify the rules for ``safe'' health advice, and use it to validate the advice. \\ \hline
  Bad code 
  & An LLM generated code that appears plausible but is functionally incorrect or references non-existent libraries and functions~\cite{code-hallu}. & Provide or generate comprehensive tests to validate the generated code. \\ \hline
  Data leakage & 
  LLMs reproduced proprietary code or personal data present in their training datasets~\cite{hallucinations}. &  Check answers for possible proprietary info (e.g., SSNs, phone \#, closed-source software licenses) and refuse to answer if info present.\\ \hline
  Security vulnerability & LLM-generated code  inadvertently introduced security flaws, such as SQL injection vulnerabilities and improper handling of user inputs. & Provide tests to expose vulnerabilities, e.g., fuzz testing for improper input handling~\cite{fuzz-testing-survey}, SQL injection attacks~\cite{sql-injection-attacks}. \\ \hline
  Legal misinformation & A lawyer submitted a ChatGPT-generate brief containing fictitious case citations leading to professional embarrassment and legal repercussions~\cite{hallucinations} & 
Check every citation in the brief against law case databases. \\ \hline
\end{tabular} \\
\vskip 0.1cm
\caption{Examples of LLM failures, their impact, and possible fixes. These fixes include better specifications: clear descriptions of what tasks should do, and clear criteria for evaluating tasks' solutions (e.g., answers). These specifications can lead to better, less ambiguous prompts and the implementation of guardrails for LLM outputs. Many LLM-based services, such as customer service chatbots, consist of multiple components, including knowledge retrieval, language understanding, and answer generation components. 
}
\label{tbl:llm-failures}
\end{table}

Table~\ref{tbl:llm-failures} provides examples of failures in today’s LLM-based systems and suggests potential fixes. Unsurprisingly, these fixes revolve around improving task specifications, underscoring their importance in building modular and reliable LLM-based systems. Indeed, we are already seeing the drive for better task specifications for LLMs, and the evidence that better specifications help. 

\begin{itemize}[noitemsep,topsep=0pt,parsep=3pt,partopsep=0pt]
    \item {\bf Prompt engineering:} 
    This involves crafting and refining prompts to achieve the desired outputs. Techniques include creating detailed prompts\cite{good-enough-prompting,prompt-engineering} (i.e., better specifications) to generate better results. Recently, DSPy~\cite{dspy} has emerged as an automated solution for prompt engineering that uses examples of desired prompts and outputs (i.e., solution specifications) along with additional hints.
    \item {\bf Constitutional 
    AI and guardrails:} 
    Constitutional AI~\cite{constitutional-ai}  establishes rules to avoid harmful outputs, enhancing task statement specifications to guide LLM answers. Guardrails~\cite{nemo-guardrails} consist of code snippets that filter prompts and outputs, either avoiding certain topics or enforcing structured formats. These can address both statement and solution specifications. 
    \item {\bf Structured Outputs:} 
    A growing number of LLMs support JSON-formatted inputs and outputs or other standardized formats~\cite{efficient-guidance, guidance, sglang}. For example, OpenAI recently introduced "structured outputs"~\cite{structured-outputs}, ensuring outputs adhere to schemas like JSON, as well as schemas described in Pydantic~\cite{pydantic} and Zod~\cite{zod}, respectively. Additionally, Trace~\cite{trace} allows users to specify a program using pseudocode and iteratively refine it. These constructs help define statement specifications (e.g., the desired output format) and, to a lesser degree, solution specifications (e.g., they typically describe the syntax of the outputs rather than their semantics).
    \item {\bf Reward models:}  
    Some of the most impressive results, such as those demonstrated by the recently released OpenAI’s o1~\cite{gpt-o1}, have been in the domains of problem-solving and reasoning—areas where it is easier to define clear task specifications. Specifically, to enhance reasoning capabilities, o1 employs reinforcement learning, which relies on an accurate reward function or model. A reward model can be thought of as a weak form of solution specification that enables the selection of the most likely correct solution among many candidates. To improve both training efficiency and scalability during testing, the reward function must be sufficiently precise to allow the system to evaluate solutions automatically.

\end{itemize}

\begin{figure}[h]
    \centering
    \includegraphics[width = \textwidth]{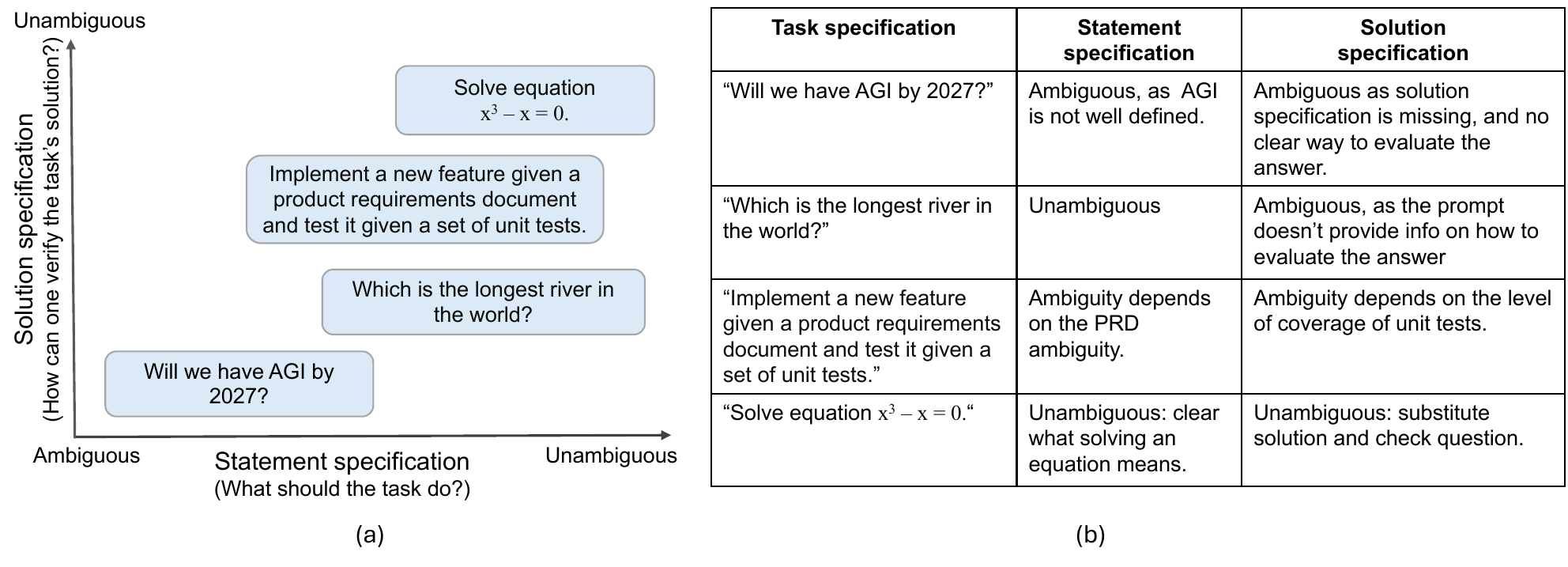}
    \caption{(a) A task specification describes (i) what the task should do (i.e., statement specification) and (ii) how should the task's solution be verified (i.e., solution specification). (b) Some tasks, such as "Will AGI happen by 2027?" are ambiguous both in terms of statement and solution specifications (e.g., AGI is not well defined, and there is no clear way to evaluate how good an answer (solution) is.), while some tasks such as "Solve $x^3 - x = 0$" are unambiguous both in terms of statement and solution specifications (e.g., one can easily verify the correctness of the solution by substituting it in the equation).
    }
    \label{fig:specs-ambiguity-examples} 
\end{figure}

\begin{figure}[h]
    \centering
    \includegraphics[width = 0.8\textwidth]{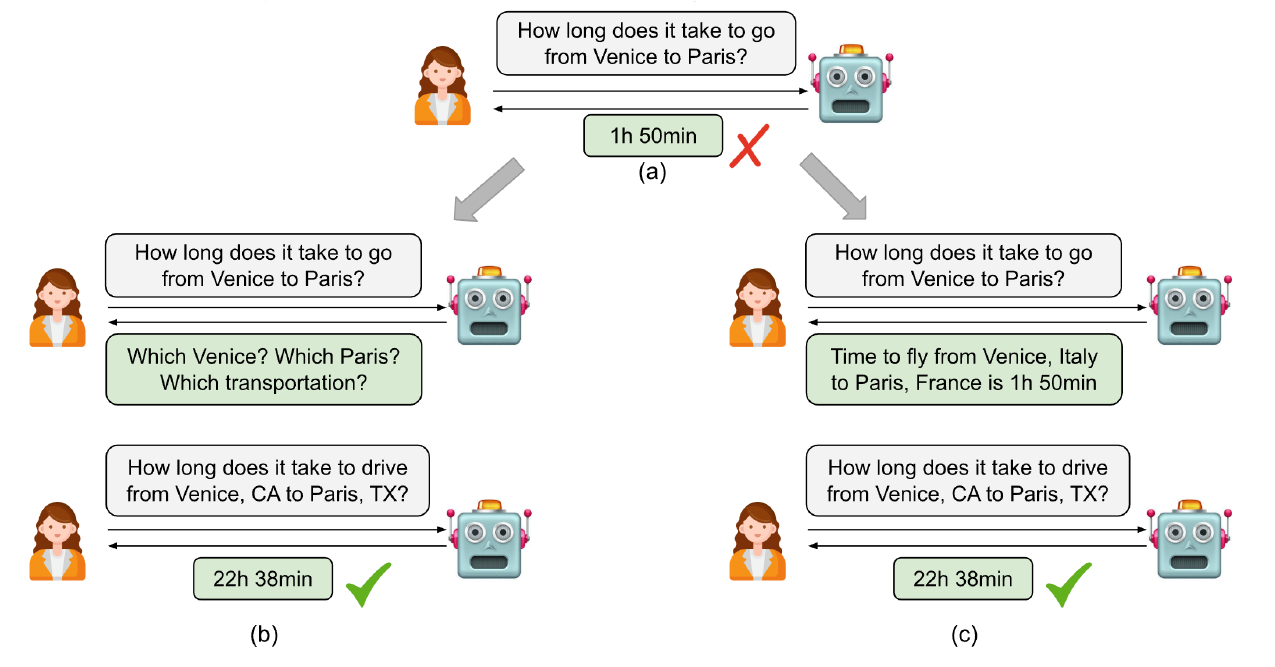}
    \caption{A simple example of prompt (statement) disambiguation. (a) The user wants to learn how long it takes to drive from Venice, CA to Paris, TX, but doesn’t mention all details in her question. As a result, the agent gives the wrong answer, i.e., the flight time from Venice, Italy to Paris, France. Two simple methods to disambiguate: (b) the agent asks for clarification about which cities and the method of transportation; (c) the agent states all the assumptions in its answer, so the user sees how to improve the prompt.
    }
    \label{fig:specification-example} 
\end{figure}

This paper advocates for building on these early efforts to develop a coherent framework for improving task specifications. 
New methodologies are essential for task specifications in this new era of AI. These include clear and measurable performance metrics, as well as approaches to ensure that desired outcomes are reliably achieved. As illustrated in Figure~\ref{fig:specs-ambiguity-examples}, LLM tasks can have specifications with widely varying degrees of ambiguity. Our goal is to move task specifications toward the top-right corner of Figure~\ref{fig:specs-ambiguity-examples}(a). The less ambiguous the statement specification, the more likely the LLM is to generate the correct output. Similarly, the less ambiguous the solution specification, the easier it becomes to verify the LLM's output.

Reducing ambiguity will enable the application of engineering principles to LLMs, fostering the development of new solutions for an expanding range of use cases. We believe these properties will help unlock the potential of compound AI systems~\cite{compund-AI} and agentic systems~\cite{agentic-systems}, which are seen by some as the next revolution in the software industry. For example, verifiability—the ability to trust LLM outputs (which requires clear solution specifications)—and the ability to make decisions without human intervention are critical challenges for today’s agentic systems~\cite{agentic-systems-challenges}. Furthermore, reducing ambiguity is valuable even for inherently ambiguous tasks (e.g., writing an email). Figure~\ref{fig:specification-example} illustrates an ambiguous prompt and two methods for disambiguating its statement specification (see Section~\ref{sec:sol-prompt-disambiguation}).

The rest of the paper is organized as follows: Section~\ref{sec:challenges} discusses the main challenges confronting today's LLMs; Section~\ref{sec:specification} examines task specifications in the context of software engineering; Section~\ref{sec:five-properties} presents the five properties enabled by these specifications, which have been instrumental in driving the growth of many engineering disciplines; Section~\ref{sec:llm-vs-sw-development} contrasts software development with LLM development; Section~\ref{sec:llms-specs} addresses the challenges of disambiguating LLM task specifications (e.g., prompts); Section~\ref{sec:prompt-dsambiguation} approaches for disambiguate task specifications; Section~\ref{sec:five-properties-llms} demonstrates how better specifications can enable the application of the five engineering properties to building LLM systems; finally, Section~\ref{sec:summary} summarizes our findings.

\section{Challenges with today’s LLMs: building modular and reliable systems}
\label{sec:challenges}
Over the past two decades, and especially in the past two years since the introduction of ChatGPT, generative AI has made enormous strides. While the rate of progress has been astounding, most of this progress has come from building larger and better models—models that often take many months, even years, to develop and require staggering computational and financial resources. Today, the cost of developing state-of-the-art models is on the order of hundreds of millions of dollars and is soon expected to climb to billions.

This state of affairs raises two challenges. First, the prohibitive costs constrain the development of these models to only a few companies. Second, the monolithic nature of these models makes it difficult to address issues when they produce incorrect outputs. Even if we have access to all the model’s parameters, it is far from trivial\footnote{While there are efforts to improve the debuggability and interpretability of LLMs~\cite{llm-survey, transformer-debugger, llm-interpretability}} to identify the root of the problem. This makes it challenging to fix hallucinations, arguably the biggest drawback of today’s LLMs. These two challenges can significantly slow the rate of progress and ultimately hamper the growth of the AI industry.

In contrast, previous engineering-driven revolutions progressed not only by improving monolithic systems but even more so by building modular systems. Take a car, for example: it is built from a myriad of components, many of which are shared across different models (e.g., wheels, brakes, engines), and these components are produced by different teams within the same company or across companies (see Figure~\ref{fig:eng-properties-example}). Similarly, an operating system is built from numerous components (e.g., scheduler, file system, memory manager), each of which is composed of smaller components (e.g., device drivers), developed by hundreds of individuals.

Decomposing complex systems into simpler ones and reusing existing components has significantly lowered the barriers to building complex systems. This approach makes it easier to fix and improve these systems by isolating faulty components and repairing or replacing them. Furthermore, modularity enables a broader set of participants to contribute to value creation, as it lowers the barrier to building new, complex systems from existing ones rather than creating them from scratch.

Perhaps not surprisingly, given these historical patterns and advantages, we have recently seen efforts to build more \emph{modular} AI systems~\cite{compund-AI, agentic-systems, is-data-all-you-need}. However, despite these efforts, breakthroughs in AI have largely continued to come from a handful of frontier labs developing larger, more powerful models.

In an effort to turn this tide, in this paper, we draw from the rich body of work on building modular  systems and apply these lessons to transitioning from monolithic to modular LLM systems. Such transformations have happened before. As engineering disciplines mature, they typically evolve from building large, monolithic artifacts to building modular ones, from using ad hoc processes to employing precise rules and standardized components.

One example is the automotive industry, which evolved from building bespoke cars—where much of the work was done by hand, and parts of the same car model were not interchangeable (e.g., a door of a Ford Model K~\cite{ford-k} wouldn’t fit another Model K car without adjustments)—to building millions of cars from thousands of standard, interchangeable components. Another example is the computer industry, which transitioned from creating very expensive, one-of-a-kind computers (e.g., ENIAC~\cite{eniac} to building affordable PCs using standardized components. These evolutions dramatically lowered the cost and barrier to building new artifacts.

In addition to building modular LLM-based systems, another significant challenge is building \emph{reliable} systems. As illustrated in Table~\ref{tbl:llm-failures}, LLM-based products and services are prone to high-profile failures across a wide variety of scenarios. To a large extent, this is due to their intrinsic flexibility: LLMs are often asked to perform ambiguous, ill-defined tasks. While embracing ambiguity has been a key to their success, deploying LLM-based products in practical scenarios often requires them to operate reliably, ideally without human supervision.

Once again, we can draw inspiration from engineering disciplines, which have developed processes and techniques to ensure reliability over time. For instance, the construction industry has implemented a variety of techniques, tests, and standards to build structures capable of withstanding earthquakes and hurricanes. Similarly, the software industry has adopted development methodologies (e.g., waterfall, agile development) and rigorous testing procedures (e.g., formal verification~\cite{tla+,coq,alloy}, fuzz testing~\cite{fuzz-testing-survey}) to create systems that can operate error-free for extended periods and control critical systems such as airplanes, cars, and nuclear power plants.

Building modular systems is closely related to building reliable systems. Reliable systems are often built as modular systems by combining smaller, reliable components. For example, a reliable car consists of a reliable engine, reliable brakes, and so on. Similarly, a reliable backend service consists of a reliable web server, a reliable database, a reliable operating system, and so forth.

In general, there are five properties which allow us to build such systems: 
\begin{itemize}[noitemsep,topsep=0pt,parsep=3pt,partopsep=0pt]
\item \textbf{Verifiability}: ability to verify wether a system works correctly.
\item \textbf{Debuggability}: The ability to debug a system if it doesn't work as expected. {\it Verifiability and debuggability are key to building reliable systems.}
\item \textbf{Modularity}: The ability to decompose a system into multiple components and combine them together.
\item \textbf{Resuability}: The ability to use existing components to build larger systems. Reusability complements modularity by leveraging existing components. {\it Modularity and reliability are key to building modular systems.}
\item \textbf{Automated decision making}: The ability to build systems that can make decisions without human supervision, which typically implies reliability.
\end{itemize}

But is there anything that ties these properties together? Our answer is "yes." Our central thesis is that the key building block enabling these five properties is \emph{specification}: the ability to describe what each component should do and how to verify its outputs. In the next section, we will illustrate how specifications are critical to enabling these properties by discussing them in the context of software engineering.

\section{Specification, the foundation of software engineering}
\label{sec:specification}
The goal of a software system is to perform a specified task. The specification provides a detailed description of what the task should achieve, including the required inputs and expected outputs. When clear from context, we will refer to “the specification of the task implemented by a system” simply as the “system’s specification.”

Ideally, specifications should be unambiguous, leaving no room for interpretation about the system’s correct behavior or output for any given input. As shown in Figure~\ref{fig:specs-ambiguity-examples}, a task specification should describe both (i) what the task should do (statement specification) and (ii) how should one verify the task's solution/outputs (i.e., \emph{solution specification}). A task specification is said to be unambiguous if both its statement and solution specifications are unambiguous; otherwise it is ambiguous. Note that these two specifications are complementary. The statement specification alone cannot guarantee the desired output, as there could be bugs in the task's implementation. Thus, we need a solution specification to be able to check wether the solution produced by the task's implementation satisfies the specification.   

Figure~\ref{fig:specs-ambiguity-examples} shows several simple examples of tasks with various degrees of ambiguity. Task "Solve equation $x^3 - x = 0$ has both unambiguous statement and solution specifications. In contrast, task "What is the longest river in the world?” has an unambiguous statement specification, but its solution specification is ambiguous, as it doesn’t provide the user with enough information to verify an answer, and it is unlikely the user has this information either. In particular, it is unlikely for a user that asks this question to know that "Mississippi", for instance, is a wrong answer.

\eat{
Putting it in another way, a task has an unambiguous statement specification if the answer to question "Does the task specification contain all information to implement it correctly?" is "yes", and an unambiguous solution specification if the answer to question "Given a task solution/output, does this solution satisfy the specification?" is "yes". These two questions are complementary. An unambiguous statement specification does not guarantee the task will provides the desired solution (e.g., due to bugs in the task's implementation), so we need an unambiguous solution specification to verify the solution indeed satisfies the specification.

Figure~\ref{fig:specs-ambiguity-examples} shows several examples of tasks with various degrees of ambiguity. Task "Solve equation $x^3 - x = 0$" has both unambiguous statement and solution specifications. In contrast, task "What is the longest river in the world?” has an unambiguous statement specification, but its solution specification is ambiguous, as it doesn't provide the user with enough information to verify an answer, and it is unlikely the user has this information either. In particular, it is unlikely for a user that asks this question to know that "Mississippi", for instance, is a wrong answer. 
}

Specifications do not exist in isolation; they typically assume a user (or another system) that can interpret them. 
For example, a user using a formal specification is expected to understand the formalism; a developer using copilot for Rust is expected to be familiar with Rust and with the code base; a user submitting a task to solve a math equation is expected to be able to check wether the solution is correct or not. Throughout this paper, unless otherwise noted, we assume an "average" user, not an expert in the intended user group. For example, in a programming task, we assume an average programmer rather than an expert; for a user asking a factual question, we assume they may not know the answer and might be misled by incorrect answers, like "Mississippi" in our "What is the longest river in the world?" example.

\eat{
Note that an unambiguous specification is not required to have a unique correct output, but it does need to provide an unambiguous solution specification to make it possible for another program (or the same program that generated the output) to verify whether the output is correct or not. An example of an unambiguous specification that admits multiple correct outputs (solutions) is “solve equation $x^3 - x = 0$”, which has solutions $\{-1, 0, 1\}$. The program producing the solution can easily verify whether the solution is correct by simply substituting it in the equation. In contrast, there are task specifications that have unique solutions, but are still ambiguous as they offer no clear solution specification. For example, consider the question “What is the longest river in the world?”. While this question has a unique answer (i.e., river Nile), there is no easy way to verify the answer unless we have access to a database of geographic facts. But if we have access to such a database, there is no need to ask the question in the first place! Thus, providing enough information to allow one check the task's output correctness with no additional information (outside the task's solution specification) is a critical property of an unambiguous task specification.
}

When it comes to software systems, there are several frameworks for writing unambiguous specifications, including {Coq}/{Gallina}~\cite{coq,gallina}, {TLA+}~\cite{tla+},  {Alloy}~\cite{alloy} and {Lean}~\cite{lean}. For example, one could use Gallina to write unambiguous statement specifications, and Coq to generate unambiguous solution specifications as proofs that the generated programs indeed satisfy their statement specifications. Unfortunately, writing such specifications is far from trivial, so these frameworks end up covering a narrow set of use cases, e.g., use cases in which the cost of failure is very high. One example is chip design where flaws can lead to delays of months and huge financial losses (see here). Another example is a reactor control system in a nuclear plant, where a failure can lead to loss of lives. 

However, most software systems do not incur such prohibitive costs on failures. These include the majority of software systems we are using daily such as text editors, chat applications, and even databases or operating systems. In those cases, a failure (e.g., crash) has relatively low cost, and it is relatively easy to recover (e.g., just restart the system). As a result, many such systems can get away with weaker forms of specifications, e.g., solution specifications that define the outputs for most common, but not all, inputs. A common example is a solution specification that leverages unit tests which don't cover all code paths. Another example is a specification that uses preconditions to filter incorrect outputs and postconditions to filter most, but not all, incorrect outputs~\cite{desgn-by-contract} (e.g., a postcondition can specify that $a > b$, but not specify precisely the relation between $a$ and $b$). 

It is important to note that an unambiguous specification does \emph{not} necessary imply a deterministic behavior, e.g., the outputs of a system are deterministic. Indeed, an unambiguous specification can describe a stochastic system, such as Markov chains~\cite{probability-stochastic} or queueing systems~\cite{queueing-systems}. For example, consider an M/M/1 queue~\cite{queueing-systems} with mean arrival rate $\lambda$ and mean service rate $\mu$. This specification is unambiguous as it leaves no room for interpretation of how the system should behave, what are the expected inputs and outputs (e.g., system utilization $\rho = \lambda/\mu$ and mean serving time $1/(\mu - \lambda)$). Of course, to verify the outputs of such systems one would need to use statistical techniques, such as confidence intervals, to determine the probability these outputs meet the systems' specifications. 

Finally, note that some stochastic systems lead to pseudo-deterministic outputs given an enough number of executions (trials), as the law of large numbers kicks in. Stochastic systems can even produce deterministic outputs. One example is using randomized search to find a solution to a problem, such as generating a program passing given tests~\cite{AlphaCode}. 

\section{The five properties of software engineering?}
\label{sec:five-properties}

\begin{figure}[h]
    \centering
    \includegraphics[width = 0.9\textwidth]{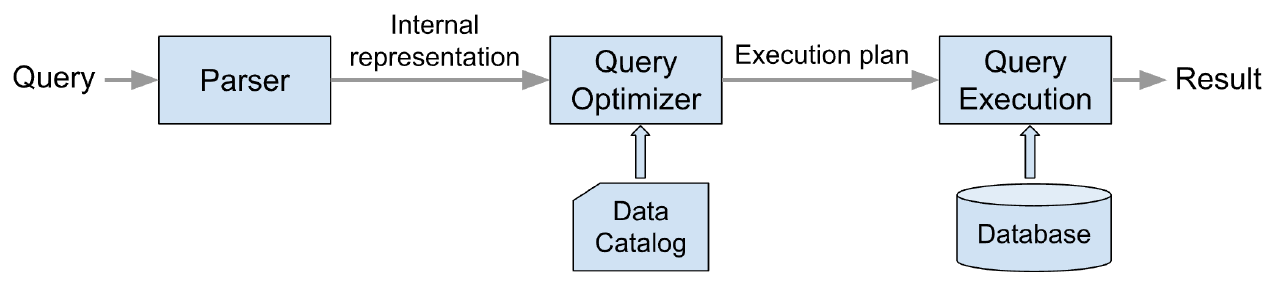}
    \caption{The main system components of a SQL query engine.}
    \label{fig:query-optimizer} 
\end{figure}

In this section, we introduce five desirable properties enabled by statement and solution specifications–verifiability, debuggability, modularity, reusability, and automated decision making–in the context of software systems. To illustrate these properties, we consider a running example using a database query engine as shown in Figure~\ref{fig:query-optimizer}. A query engine consists of several components, (1) a parser which takes a query and produces an intermediate representation of the query such as an abstract syntactic tree (AST), (2) a query optimizer that takes the AST as input and applies various optimization techniques to generate an efficient query execution plan, and (3) a query execution component that executes the query plan to produce the final result. In addition, the query engine interacts with a data catalog that maintains metadata information (e.g., table schemas, location) needed by the query optimizer to produce the execution plan, and the database storing the actual data. 

As illustrated in Figure~\ref{fig:query-optimizer}, one of the properties of the query engine is \textbf{modularity}, i.e., the query engine consists of multiple software components. This modularity enables us to build these components in parallel and then assemble them together. A second related property is \textbf{reusability} which is the ability to use existing components when available. For example, instead of writing a query parser from scratch, we can use an existing library such as {BlazingSQL}~\cite{blazing-sql} or {JSQLParser}~\cite{jsql-parser}, to speed up building the query engine. 

Modularity and reusability are enabled by the fact that each component has a clear \textbf{specification} which, again, consists of the statement specification (e.g., PRD) and the solution specification (e.g., unit and integration tests). These specifications help ensure that the output of a component matches the expected input of the next component, and help us reason about the end-to-end behavior of the system. For instance, we expect the query optimizer to affect only the query performance, not the outputs produced by the query which should remain the same.

While modularity and reusability can help building new systems from existing components, in many cases we need to build components from scratch. Implementing a new component requires two other properties: \textbf{verifiability} and \textbf{debuggability}. Verifiability is enabled by the solution specification which describes the correct outputs of the task implemented by the component.
In most cases the solution specification consists of a set of input-output tests (e.g., unit tests, integration tests, statistical tests, etc).
Alternatively, it can come in the form of preconditions and postconditions which filter out invalid inputs and outputs~\cite{desgn-by-contract,eiffel}.
In our query optimizer example, a verifier would check that none of the optimizations (e.g., push down predicates, join reordering) change the query result. 

Debuggability is the process of fixing the implementation if the verification fails. This involves locating the bug and fixing it. Debuggability is enabled by the hierarchical modularity of software systems. In our example, we can start with the query engine, and identify which component is buggy. Furthermore, within that component, we  identify the module, function, and finally the sequence of instructions causing the bug. Then, we fix the bug. This top-down process of locating the bug is enabled by the fact that each component has  a clear solution specification. 

A final property of software systems is automated \textbf{decision-making}. This enables a system to make decisions without human supervision. This property is key to addressing use cases where humans are too slow or error-prone. One simple example is the query optimizer in our SQL engine, which computes an efficient query plan automatically. Other examples include mission-critical systems such as anti-lock brake systems for cars, automatic takeoff and landing systems for airplanes, and reactor control systems in nuclear power plants.

Unlike automated decision-making systems, the majority of existing software systems (e.g., chat apps, IDEs, games, etc.) assume a human in the loop. As such, these systems typically rely on humans to detect failures and provide remediations (e.g., restarting a program when it crashes or hangs). In contrast, automated decision-making systems do not rely on humans, which means they need to avoid failures in the first place. As a result, these systems tend to have less ambiguous (ideally, unambiguous) statement and solution specifications. 
Indeed, automotive and aerospace industries often use formal tools for specification and verification, such as {SPARK}~\cite{spark} and the {SCADE Suite}~\cite{scade}. 

\section{Comparing LLM development and software development}
\label{sec:llm-vs-sw-development}

The recent progress of LLMs has sparked the imagination of many with Andrej Karpathy famously stating that {English is the hottest new programming language}''~\cite{programming-language-english}, which suggests that one can ``program'' LLMs by simply describing tasks in natural language. 

\begin{figure}[h]
    \centering
    \includegraphics[width = 0.9\textwidth]{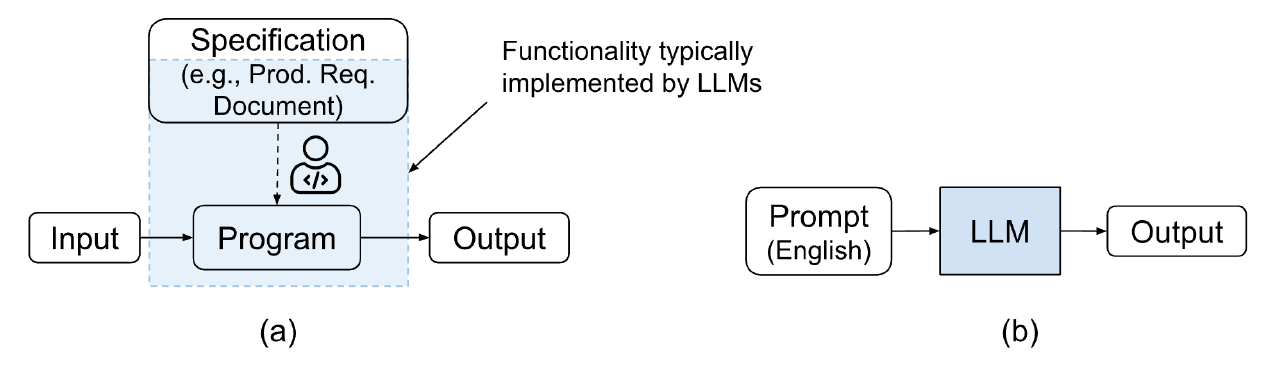}
    \caption{High-level architecture of (a) developing a program for a given task vs (b) using an LLM to accomplish the same task.The LLM obviates the need for writing the program and doesn’t need a detailed specification. The LLM prompt contains both the description (specification) of the task and its input.}
    \label{fig:swe-vs-llm} 
\end{figure}

Figure~\ref{fig:swe-vs-llm} contrasts (a) developing a program for a given task using traditional software engineering with (b) accomplishing the same task using LLMs. With software engineering, a developer takes a task's statement specification and writes a program to implement that task. The program then takes an input and generates an output which can be verified using the solution specification. 
In contrast, an LLM takes a description (e.g., a high-level statement specification) of the task together with the task’s input as a prompt and produces an output. Thus, LLMs have the potential to obviate the need for writing programs and for detailed specifications. When it comes to specifications, LLMs typically trade unambiguity for the ease of writing these specifications in natural language,

\begin{figure}[h]
    \centering
    \includegraphics[width = \textwidth]{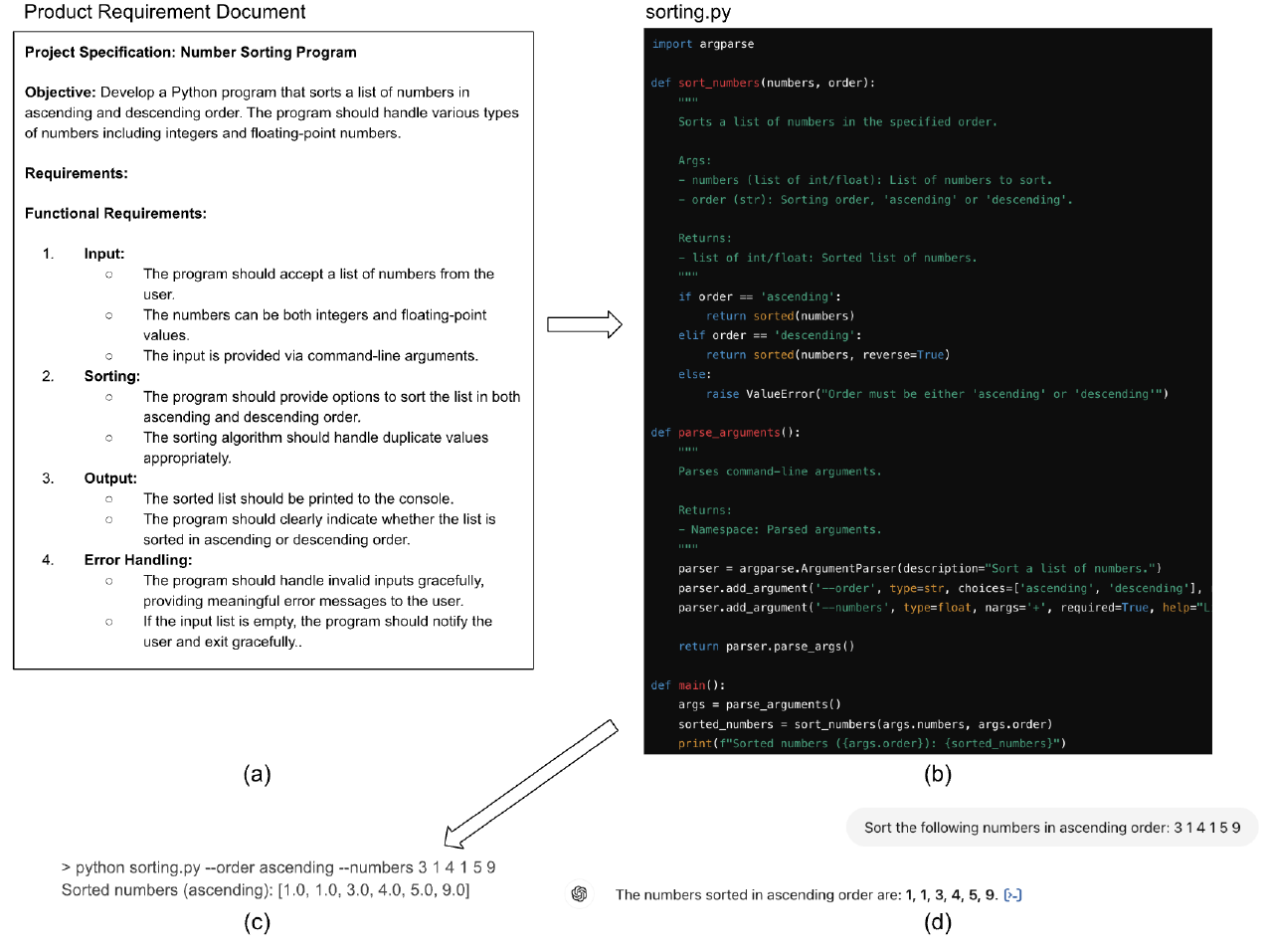}
    \caption{The task of ordering an array of numbers. (a) Specification; (b) Program implementation; (c) Executing the program on an example input. (d) A prompt asking ChatGPT to order the same example input.}
    \label{fig:sort-example} 
\end{figure}

Figure~\ref{fig:sort-example} shows a simple task to sort a list of integers, implemented by using both a program and an LLM. Figure~\ref{fig:sort-example}(a) shows the statement specification in the form of a PRD, Figure~\ref{fig:sort-example}(b) shows a Python program implementing the PRD, and Figure~\ref{fig:sort-example}(c) shows a call to the program with input list: 3, 1, 4, 1, 5, 9. In contrast, Figure~\ref{fig:sort-example}(d) shows an equivalent LLM solution consisting of a prompt that contains both a high-level statement specification of the task (i.e., “Sort the following numbers in ascending order”) and the same input. Note that none of these cases shows a solution specification.

While the example in Figure~\ref{fig:sort-example} works well and shows the full power of LLMs, unfortunately, in many cases the prompts are ambiguous and users don’t get the desired answers. We discuss this challenge and approaches to address it next.

\section{Task specifications in LLMs}
\label{sec:llms-specs}

When it comes to LLMs, we are faced with a conflict between (1) the accessibility and generality of the natural language (i.e., which allows anyone to specify any task), on one hand, and (2) its inherent ambiguity, on the other hand. Next, we examine approaches on how to resolve this conflict by making it easier to write clear (unambiguous) specifications for a growing number of LLM tasks. This will allow us to leverage engineering principles to implement these tasks and provide solutions for an increasing range of applications. 

\subsection{Are LLM prompts inherently ambiguous?}
\label{sec:llm-prompt-ambiguity}

With LLMs, we specify tasks using prompts. We say that a prompt is inherently ambiguous if it is very hard, or infeasible, to disambiguate it. As before, we need to differentiate between statement and solution specifications. An example of a prompt that is ambiguous both in terms of statement and solution is “Write a poem about a white horse in Shakespeare’s style”. Such prompt is inherently ambiguous as there is no clear way to disambiguate it. Indeed, the user might not even know what a good poem looks like before seeing one!  

On the other hand, consider the prompt “How long does it take to go from Venice to Paris?”. This prompt has an ambiguous statement specification since it is not clear which “Venice'' and which “Paris” the user is referring to, i.e., is it the “Venice” in Italy or the one near Los Angeles? Is it the “Paris” in France or the one in Texas? Does the user intend to travel by car, train, or airplane? This prompt's statement specification is not inherently ambiguous as the user can easily disambiguate it by adding the missing information, as shown in Figure~\ref{fig:specification-example}(b). However, note that the solution specification of this prompt is still ambiguous as it is not clear how to check whether the answer is correct from the information provided by the prompt alone. 

\begin{figure}[h]
    \centering
    \includegraphics[width = \textwidth]{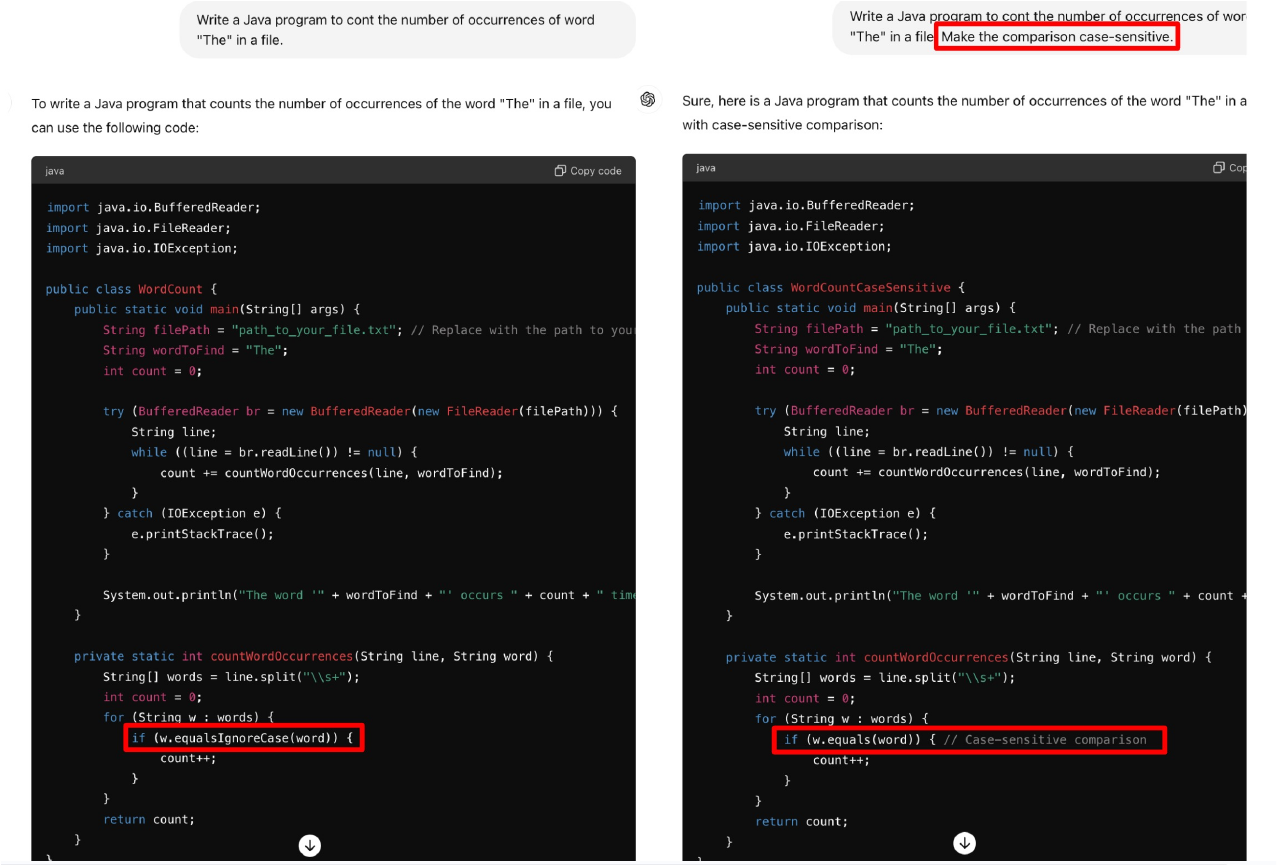}
    \caption{Consider the prompt, “Write a Java program that counts the number of occurrences of word “The” in a file”. (a) The generated program uses case-insensitive matching despite the fact that the user capitalized the first letter of “The” which might indicate she intended case-sensitive matching, (b) 
Eliminating the ambiguity by adding “Make the comparison case-sensitive” leads to the desired result.}
    \label{fig:word-search-example} 
\end{figure}

In another example (see Figure~\ref{fig:word-search-example}(a)), consider the prompt, “Write a Java program that counts the number of occurrences of the word “The” in a file. Similarly, this prompt has an ambiguous statement specification as it doesn’t say whether the match should be case sensitive or not (though in this case one could argue that the user’s intent was to have a case-sensitive match since she capitalized the first letter in “The”). Again, the prompt's statement is not inherently ambiguous as a user can easily disambiguate it by adding “Make the comparison case-sensitive”, and, by doing so, she gets the desired result (see Figure~\ref{fig:word-search-example}(b)). At the same time, this prompt doesn't provide a solution specification either. However, one could add such specification by providing some unit tests, e.g., provide files with a known number of "The" words and extend the prompt to provide the expected number of "The" words for each test file.

\section{Disambiguating LLM specifications (prompts)}
\label{sec:prompt-dsambiguation}

As mentioned above, while most of LLM prompts are ambiguous, in many cases this ambiguity is not fundamental; it is merely a reflection of the fact that writing unambiguous specifications is often hard. In this section, we suggest several approaches to make it easier to disambiguate LLM prompts. We do so by drawing our inspiration from how humans deal with ambiguity when they communicate with each other. 

\subsection{How do humans deal with ambiguity?}
\label{sec:human-ambiguity}

People use natural language every day to describe tasks for others to perform, such as "Buy one gallon of milk," "Bring one more chair to the table," or "Let’s meet at the market hall at 7" (see Table~\ref{tbl:human-disambiguation-examples}). These examples provide both a statement specification (i.e., what to do) and a solution specification (i.e., enough information to verify whether the task was performed correctly). For example, one can check that "the milk was indeed bought," "a new chair was brought to the table," and "the person attended the meeting."

However, since natural language is inherently ambiguous, the tasks it describes are often ambiguous as well, including the examples above. So, how do people handle this ambiguity? They typically rely on three main approaches.

First, most people's interactions are \emph{multi-turn}, e.g., dialogs, discussions, etc. This enables people to refine their task’s description if ambiguous. If Alice asks Bill to do something, and Bill is confused, he will ask for clarifications. This will give Alice the opportunity to clarify the task’s description, and Bill the opportunity to build more context about the task. This pattern is very common. Table~\ref{tbl:human-disambiguation-examples} shows some examples of clarification questions: "What kind of milk? What brand? By when do you need it?", or "Which chair?", or "Which market hall? Which stall in the market hall? 7am or 7pm?"


Second, people assume a \emph{task-aware context}. Such context is built from performing similar tasks in the past or extra information related to the given task. Referring again to the tasks in Table~\ref{tbl:human-disambiguation-examples}, if this is not the first time you are buying milk, you probably already know which kind of milk or which brand you need to buy, so no need to ask. Also, if you know that the milk is needed for baking a cake for dinner, you might already know by when you need to buy the milk.

Third, people often “model” other people they communicate with. They do so by learning from the repeated interactions with these people, what they like and don’t like, what is the most effective way to ask these people to perform a task, etc. We call this, \emph{people-aware context}. In the example, "Let's meet at the market hall at 7", if you know the person asking you this is a late morning person, you can easily infer this must be 7pm not 7am. 

Interestingly, people leverage similar approaches when interacting with a chatbot like ChatGPT as well: they often use multi-turn interactions to refine their asks, assume the prior context from the conversation is available when asking the next question, and learn what the LLMs are good at or they are not got at (e.g., what kind of questions are more likely to produce hallucinations).

In summary, humans typically communicate iteratively by assuming some shared context. As such, they often leave details out of the conversation, as these details can be easily filled in via subsequent questions or can be inferred from shared context. Natural language has evolved over thousands of years to let humans communicate with each other efficiently. However, this does not necessarily make it the most effective way to communicate between an LLM and a human, two LLMs, or an LLM and an existing software system (e.g., function calling). Thus, it should come as no surprise that we need new techniques to improve communication in those cases.

\begin{table}[]
\begin{tabular}{|p{2.4cm}|p{3cm}|p{5.5cm}|p{2.8cm}|}
\hline
\textbf{Task} &
  \textbf{Multi-turn} &
  \textbf{Task-aware context} & 
  \textbf{People-aware context} \\ \hline
"Buy a gallon of milk." &
   "What kind of milk?" \newline
   "Which brand?" \newline
   "By when?" &
   I am buying milk every week, so 
   I already know the kind of milk and brand.\newline 
   I know you need the milk to bake a cake 
   so I know when you need the milk by. &
   I know you like organic, reduced-fat milk. \\ \hline 
"Bring a chair to the table." &
   Which chair? &
   I already know which chair to bring, as this is not the first time you asked me to do it. &
   I know you like firm chairs. \\ \hline
"Let's meet at the market hall at 7" &
   Which market hall? \newline
   Which stall? \newline
   7am or 7pm? &
   This is a regular meeting point and it's the usual time we are meeting. &
   I know you are a late morning person, so it's likely 7pm. \\ \hline
\end{tabular} \\
\caption{Example of everyday tasks, and approaches people use to disambiguate them.}
\label{tbl:human-disambiguation-examples}
\end{table}

\subsection{Solutions for prompt disambiguation}
\label{sec:sol-prompt-disambiguation}

In this section, we discuss several techniques to disambiguate LLM prompts drawing inspiration from how humans disambiguate their communication. Most of the techniques here focus on tasks' \emph{statement} specifications.

\textbf{Iterative disambiguation:} As mentioned, in real life, people often ask questions to clarify the description of a task provided by another person. One example is a product manager refining a product requirement document (PRD) based on the feedback from a developer: when a developer is confused about the PRD, she asks questions, which typically lead to the program manager adding clarifications or new information to the PRD with the goal of disambiguating it. 

One natural approach would be to mimic this process by having an LLM ask a human questions to help her disambiguate her prompt (see Figure~\ref{fig:specification-example}(b)). But how can an LLM figure out whether a prompt is ambiguous? One possibility is to leverage the observation that an ambiguous prompt is more likely to generate different outputs than an unambiguous one.\footnote{While in many cases a prompt with different answers indicates ambiguity, this is not always the case. An unambiguous prompt can have different answers, all correct, e.g., "fastest driving route from A to B" can return two different routes that take the same time.} In particular, if an LLM generates multiple outputs for the same ambiguous prompt, the chances are some of these outputs will be different. Consider again our example in which we want to generate a program that counts the number of occurrences of the word “The” in a particular document. By generating many programs~\footnote{To increase the chance of generating different programs for an ambiguous prompt, one could increase the temperature, use different LLMs, use different languages, rewrite the prompt, etc.} we might end up with programs that provide different outputs depending on whether the comparison is case sensitive or not. By showing the differences between the programs to the user, we can help her realize that she needs to add the case sensitivity matching requirement to the prompt. 

\begin{figure}[h]
    \centering
    \includegraphics[width = \textwidth]{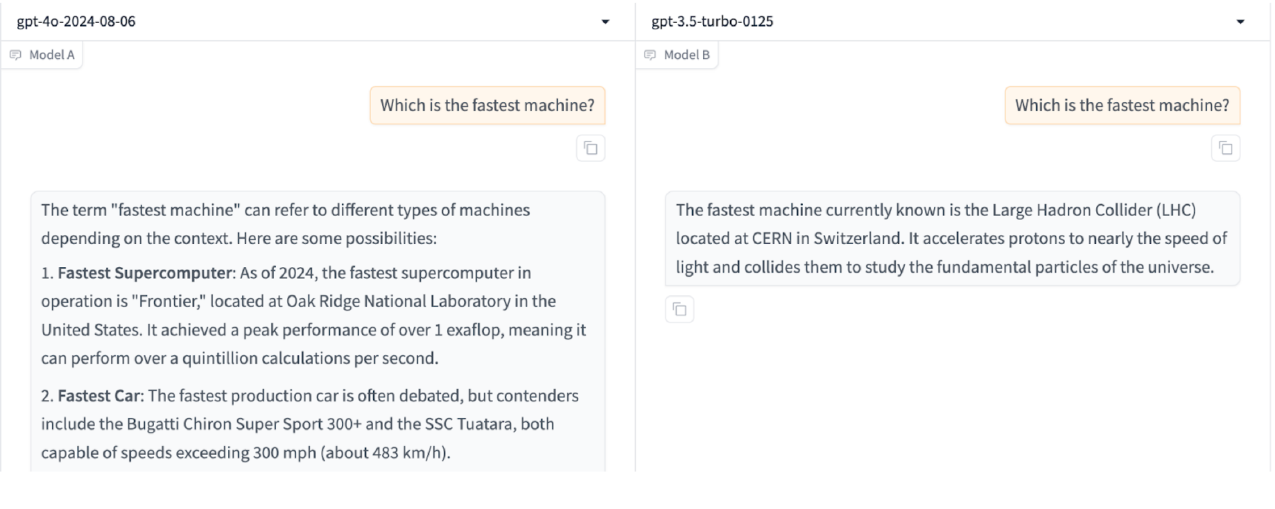}
    \caption{Two LLMs give two different answers to the prompt:  “Which is the fastest machine?”. This suggests the prompt is ambiguous and the user should disambiguate it, e.g., “What is the fastest car?”. }
    \label{fig:fastest-machine-example} 
\end{figure}

A more advanced approach would not only show the differences between programs to the user, but also suggest how to disambiguate the prompt, e.g., “Consider specifying whether the matching is case-sensitive or not?”. In another example, consider prompt “Which is the fastest machine?”. This prompt is also ambiguous. Is it the fastest computer? The fastest car? And, if it is the fastest car, is it the fastest road car? The fastest race car? Or maybe the fastest vehicle period, i.e., the kind of cars using jet engines that are purposely built for land speed records? Again, one way to disambiguate the original prompt is to use one or more LLMs to provide different outputs that expose the ambiguity (see Figure~\ref{fig:fastest-machine-example}). 

Yet another approach would be for the LLM to give answers that clearly state the assumptions it made in response to ambiguous prompts. For example, given the question “How long does it take to go from Venice to Paris?”, an LLM might answer “The \emph{flying} time between Venice, \emph{Italy} and Paris, \emph{France} is 1h and 50min”, (The words in italic capture the assumptions made by LLM in answering the question.) This will make it easy for the user to check whether this answers the question she had in mind and, if not, improve the prompt specification, e.g. “How long does it take to drive between Venice, CA to Paris, TX?” (see Figure~\ref{fig:specification-example}(c)). 

Note the last two approaches might require finetuning or post-training an LLM to either provide suggestions to disambiguate prompts or state the assumptions for its outputs to fill the gaps in the prompts’ specifications. 

There is already an active area of related research in programming languages to identify the user’s intent~\cite{interactive-codegeneration-test-driven,llm-transform, llm-driven-user-intent}. This research can be a promising base to build on.

\textbf{Domain-specific rules:} People in the same profession or organization share context in the form of predefined rules or special semantics associated with words or phrases. For example, when a soldier hears from his superior that she should perform a task at 11:00, she knows this is 11am (not 11pm) since the military uses the 24h time format to tell the time. In another example, for software developers, words like “class”, “loop”, “variable” have very precise meanings; the same words are ambiguous in a colloquial conversation. Finally, a programming language adds another layer of syntax and semantics which further helps to disambiguate the meaning of statements written in that language. 

In domains where there are strong rules or operating procedures we can use those rules and procedures to disambiguate the specifications. Examples are the civil or penal codes, standard-operating-procedure manuals in hospitals, labs, etc. Note that new rules are already being proposed specifically for LLMs, such as {Constitutional AI}~\cite{constitutional-ai}. Constitutional AI is a list of rules to align LLMs so that to minimize the harm they can cause. Some areas of research are (1) capturing these rules effectively in the prompt (e.g., select only the rules relevant to the task), (2) enforcing/verifying that these rules were followed by the output, i.e., solution specification. The latter is particularly challenging as most of these rules are not formal. Thus, formalizing these rules when possible is another area or research. 

Finally, another important area of research is defining new domain-specific rules. One example is capturing the laws and constraints in the physical world, e.g., people do not walk upside down unless they are in space, they do not walk through walls, etc. This is related to the ongoing work on world models~\cite{autonomous-machine-intelligence, world-models}.

\textbf{Monitor, learn \& build context:} Humans often create models (context) about the external world, including people they are interacting with. For example, we communicate and explain things differently to children, to our friends, to colleagues, or to our family. In some of these cases, we do so by learning what is the most effective way to communicate with a particular person after repeatedly interacting with that person (see Section Section~\ref{sec:human-ambiguity}). One area of research is to identify the prompts a model is good at (e.g., it excels in generating python programs for data processing) and not good at (e.g., it does a mediocre job on world problems). Also, we can learn the style of the prompt that maximizes the accuracy of model's answer. Based on these learnings, we can change prompts or route prompts to the best model to handle it. 

\textbf{``The stranger test'':} If there is no shared context available when prompting an LLM, one approach is to provide as much as possible information in the prompt, e.g.,  just assume the prompt is intended for a stranger. To illustrate this approach, consider a text-to-image task. Even though such a task is inherently ambiguous, we can still disambiguate it substantially by thinking of the prompt as a one-shot specification for a stranger. Consider you want to generate an image of the “Golden Gate Bridge at sunset” (see Figure~\ref{fig:hallucinations}(c)).  Then you should provide as much information as you can to increase the probability you get the picture you’d likely like from someone who knows nothing about your artistic preferences, e.g., the picture should use pastel colors, there should be realistic proportions between the bridge and buildings, the sun should set opposite to the city, etc. Basically, you will try to externalize the specification you have in mind about how the picture should look like.  

Another way to externalize such a specification is by showing the user various pictures and asking her to say which picture she likes and which she does not. Based on this, we can infer some user’s preferences (e.g., pastel colors vs bright colors, photograph-like pictures vs cartoon-like pictures, etc) and use these preferences to enhance the prompt. Note this is another way of building context. 

\textbf{Structured outputs:} In an effort to improve reliability, many LLMs have started to provide  support for describing their outputs and inputs in JSON format\cite{efficient-guidance}. Furthermore, OpenAI has  recently introduced "structured outputs"~\cite{structured-outputs}, a feature that not only ensures the output is in JSON format but that it adheres to a particular JSON schema.
Furthermore, Trace~\cite{trace} enables users to use pseudocode to specify a program and then use an iterative process to generate and refine the program. These constructs are one step closer to providing statement and solution specifications for programming tasks. In particular, a pseudocode provides a statement specification (i.e., it specifies what the task should do) while a schema provides a weak form of solution specification as it can be used to verify whether the output matches the given schema. However, the pseudocode can be still ambiguous, and an output schema captures mostly the syntax and not the output's semantics. As such, most of the disambiguation techniques discussed in this section apply to structured outputs as well. An interesting  research direction is taking advantage of the additional structure and application domain (e.g., program synthesis) to develop more specific disambiguation techniques. Such techniques could borrow from and extend traditional formal specification research~\cite{formal-specification-survey,formal-specification-testing}. Finally, while providing the pseudocde can help providing better solutions, writing such a pseudocode is not easy, especially for complex programs. One research direction here is to find the right balance between the level of detail of the pseudocode and the correctness of the generated program.  

\textbf{Instruction following:} Even if a prompt is unambiguous, it is of little use if the LLM cannot provide a solution for it. As such, the ability of the LLM to follow the prompt’s specification is critical to eventually find a solution. Thus, a future direction is to improve an LLM ability to follow a prompt’s specification, an already active area of research. This might involve not only test-time compute, but also pre-training and post-training.

\section{The five engineering properties in the LLM context}
\label{sec:five-properties-llms}

In this section, we show how we can leverage the five engineering properties in the context of LLMs, and in the process identify some key challenges and suggest several approaches to address them.

\subsection{Verifiability: challenges and solutions}
\label{sec:verifiability}

Verifiability refers to the ability to assess whether the task's implementation adheres to the statement specification (e.g., a program's PRD). This reduces to checking that the task's outputs satisfy the \emph{solution} specification (e.g., unit tests). Consequently, the verification depends strongly on the solution specification. If the solution specification is ambiguous, verification may fail.

\begin{figure}[h]
    \centering
    \includegraphics[width = \textwidth]{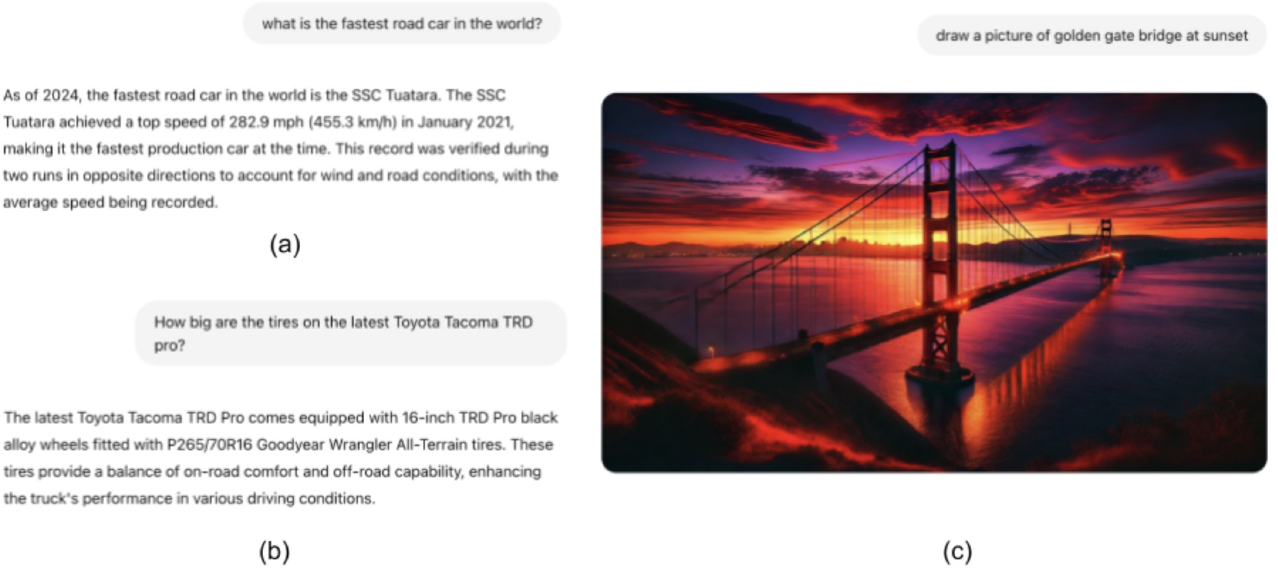}
    \caption{Example of prompts without clear solution specifications: (a) SSC Tuatara is not the fastest road car in the word; Koenigsegg Jesko Absolut is (330 MPH); (b) Toyota Tacoma TRD pro doesn’t use 16” wheels; it uses 18” wheels; (c) the picture doesn’t show Golden Gate Bridge at sunset, as the sun sets on the opposite side of the city.}
    \label{fig:hallucinations} 
\end{figure}

Figure~\ref{fig:hallucinations} illustrates examples of prompts lacking clear solution specifications. In such cases, a user will likely fail to assess the correctness of the output: (a) the user may not know whether a faster car exists than the one mentioned in the response; (b) the user may be unaware of the size of the Toyota Tacoma's wheels; and (c) the user may not know where the sun sets in San Francisco. Note these are examples of "hallucinations", i.e., outputs that look correct but are not.

\begin{figure}[h]
    \centering
    \includegraphics[width = \textwidth]{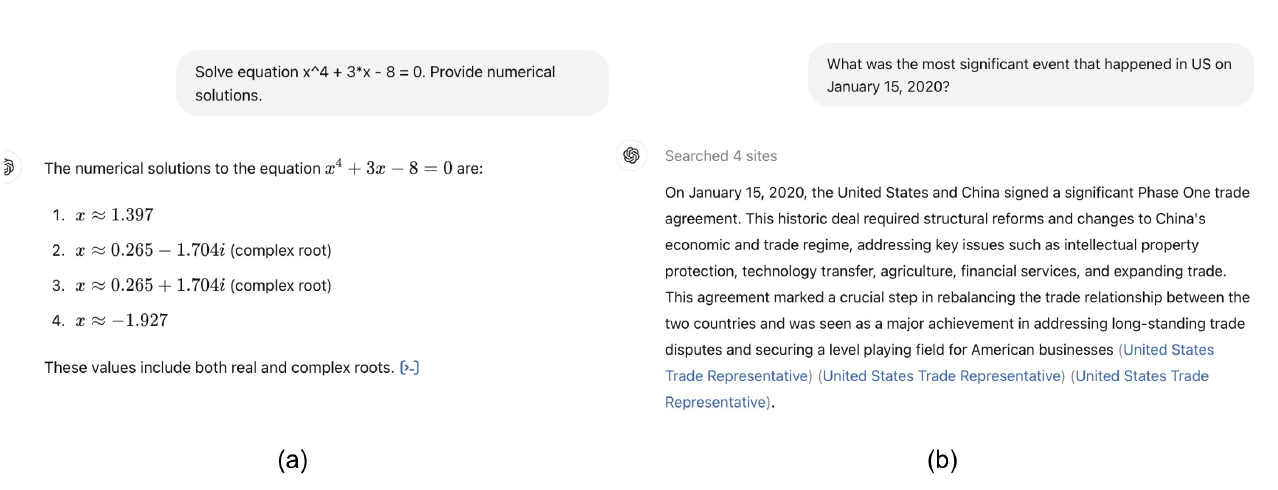}
    \caption{One can check the results provided by ChatGPT using (a) a calculator, and (b) taking a look at the documents linked in the results. }
    \label{fig:equation-example} 
\end{figure}

Even when a task has a clear solution specification, assessing whether its implementation works as intended is not always straightforward. This often requires additional context or tools. For instance, if we prompt an LLM to solve the equation "$x^4 + 3x - 8 = 0$" (note that here the equation serves as both the statement and the solution specification), a user may have to use a calculator to verify that the generated solution is correctnes (see Figure~\ref{fig:equation-example}(a)). In another example, an LLM might generate references to reputable documents, which the user may have to cross-check to verify the accuracy of the answer (see Figure~\ref{fig:equation-example}(b)).

Next, we propose several directions for future work aimed at improving solution specifications, as opposed to the statement specifications discussed in Section~\ref{sec:sol-prompt-disambiguation}. Additionally, we suggest exploring the use of external tools to verify a task's implementation effectively.

\textbf{Proof-carrying-outputs}: An unambiguous statement specification is usually insufficient to determine whether an implementation truly satisfies the specification. We also need a solution specification. Consider a code generation task with a formal (unambiguous) statement specification. How can we ensure that the generated code satisfies the specification for every possible input? One approach is to generate not only the code itself but also a proof that the code adheres to the specification, leveraging techniques such as \emph{proof-carrying code}~\cite{pcc-necula97}. In this case, the proof serves as an unambiguous solution specification.

\textbf{Step-by-step verification}: A large body of work has demonstrated that asking the LLMs to provide step-by-step solutions (e.g., {chain-of-thoughts}~\cite{chain-of-thoughts}, {tree-of-thoughts}~\cite{tree-of-thoughts}, {self-notes}~\cite{self-notes}, {scratchpad}~\cite{scratchbad}) increases the accuracy of the solutions. Fortunately, this also makes the verification easier, as we can independently verify every step of the solution, and in many cases each of these steps is easier to verify than the end-to-end solution. Of course, each step being correct doesn’t guarantee end-to-end correctness (e.g., there might be missing steps, or steps that are correct in isolation but inconsistent across them). However, early evidence shows that step-by-step verification can improve the verification accuracy, e.g., Process Supervision~\cite{process-supervision}. 

\textbf{Execute-then-verify}: Consider an output containing instructions. In many cases, it is easier to verify whether the output satisfies the task's solution specification by executing the instructions and observing the results. For example, it is often easier to verify the correctness of a program by running it and checking its outputs, than by analyzing the code alone. However, this approach poses a significant challenge: if the result is incorrect, it can lead to irreversible mistakes. For instance, imagine a program mistakenly deleting important files.
To mitigate this risk, it is essential to ensure that execution is either reversible (i.e., can be undone) or does not alter the original state of the system. For reversibility, one approach is to checkpoint the system state before each operation. Another is to log each operation along with its corresponding undo operation. For non-altering execution, a viable strategy is to run the program in an isolated copy of the original environment, such as within a virtual machine. One project exploring these approaches is GoEx~\cite{GoEx}.

\textbf{Preconditions \& postconditions}: 
One technique we can borrow from programming languages to disambiguate a task's solution specification is the use of preconditions for the task’s input and postconditions for its output~\cite{desgn-by-contract, eiffel}. For instance, when asking which is the fastest car, if the user knows that cars with a maximum speed exceeding 300 mph exist, they could specify this as a postcondition. Another promising direction is to leverage LLMs to generate postconditions directly from the prompt. For example, if the question is, “What is the tallest building west of the Mississippi?”, the LLM might generate a postcondition requiring a check that the city where the building is located is within the United States and that its geographic coordinates are indeed west of the Mississippi River.

\textbf{Statistical verification:} While so far we have primarily focused on verifying whether a query answer satisfies its specification, in some cases, the focus shifts to the “aggregate performance” of an LLM-based system. This includes metrics such as its score on benchmarks like MMLU~\cite{mmlu} or performance in real-world scenarios using platforms like ChatBot Arena~\cite{chatbot-arena}. A particular challenge in the latter setting is the lack of control over the workload (i.e., users’ queries). Changes in the workload can significantly impact the system’s performance. One approach to address this is to measure the LLM system’s performance at deployment and continuously monitor it to detect any degradation over time. In this case, the performance metric, as measured at the deployment time, becomes the solution specification. A promising research direction is to develop statistically robust techniques for evaluating system performance in real-world conditions and detecting performance changes quickly and effectively.

\subsection{Debuggability: challenges and solutions}
\label{sec:debuggability}

When the verification of a task's output concludes that the output is incorrect, the next step is to debug (fix) it. Unfortunately, debugging an LLM-based system is inherently challenging. Unlike a traditional program, an LLM is virtually a black box. Even with full access to its weights, it is extremely difficult to correlate individual weight values with the semantics of the output. The traditional approach to debugging LLMs involves trial and error. This typically includes rewriting the prompt (prompt engineering) to influence the output or leveraging the stochastic nature of LLMs to generate alternative outputs for the same prompt. While these techniques can often improve output quality, they are neither straightforward to apply nor do they guarantee correctness.
Debugging an LLM-based system is akin to debugging a huge program by looking at the output and changing a few lines of code and hoping for the best, without using any debugging tools, not even printfs!

Below, we propose several techniques for correcting the outputs of an LLM task, assuming the task has a clear specification. These techniques involve generating multiple prompts based on the original prompt and, in some cases, combining the outputs of these prompts to achieve the desired results.

\textbf{Self-consistency}: One research direction is to generate multiple outputs for the same task, outputs that are expected to be consistent, and then check whether they are indeed consistent. Assume a task with a statement specification, but no solution specification. One possibility would be to use the statement specification to generate not only (1) the task's solution, but also (2) the solution specification, and (3) then use the solution specification to check the actual task's solution. For example, given a PRD, we can generate both the program and the unit tests and check whether the generated program passes the unit tests. If yes, we conclude that there is a good chance the program is correct. Here we assume that there is a relatively low probability that both the generated program and the unit tests are incorrect, and the incorrect program passes the incorrect unite tests. Examples of works that already leverage this technique are {AlphaCodium}~\cite{AlphaCodium} and {AlphaCode}~\cite{AlphaCode}.

\textbf{Mixture of outputs}: A related and complementary technique is to generate many outputs for the same task and then select the best output, or derive a better answer than any output that has been generated so far. Examples are using majority voting, or another LLM acting as a “judge” to select the best (or the most correct) output~\cite{llm-as-a-judge,network-of-networks}, or even to aggregate these outputs to obtain a better one~\cite{network-of-networks,llm-juries}. 

\textbf{Process supervision}: This is a recently proposed {approach}~\cite{process-supervision}, and it is related to step-by-step verification. If a step fails during step-by-step verification, we can ask the same LLM or another LLM to fix the solution based on this feedback. We can employ this approach iteratively until there are no longer errors in the output. If we cannot fix the output (maybe after a predefined number of iterations) we provide the feedback to the user. 

\textbf{Automatic prompt generation}: Most existing prompt engineering techniques are manual. However, given the huge number of possible prompts describing the same task, one possible direction is to generate these prompts automatically until we get the desired output. One project following this approach is DSPy~\cite{dspy}.

\textbf{Divide and conquer}: As mentioned above, one reason LLMs are difficult to debug is because they are black boxes. One possibility to alleviate this challenge is to decompose a task into subtasks where each subtask implements a narrow functionality by eventually calling an LLM (which in turn might involve a function call). The intuition here is that it is easier to verify and debug a specialized sub-task than a general task. For example, assume that you want to book a vacation to Rome. Then, it is easier to verify a plane ticket reservation than the entire trip. Note this is related to planning and {compound AI} systems~\cite{compund-AI}, where an LLM provides a plan and each step in the plan is implemented by another LLM call. Some of these LLM calls can implement function calling. This is also related to the recently proposed "structured outputs"~\cite{structured-outputs} in which a developer can provide the pseudocode of a task which consists of multiple functions. This helps one verify and debug the implementation of each of these functions in isolation.

\textbf{Statistical debugging:} LLM-based systems often consist of multiple components~\cite{compund-AI,agentic-systems}, such as knowledge retrieval, language understanding, solution generation, and solution selection. As discussed in Section~\ref{sec:verifiability} (see "Statistical Verification"), in some cases, we are interested in the system's overall performance rather than the correctness of individual answers—for instance, in a customer support system, we may prioritize overall user satisfaction. How can we debug such a system? How can we pinpoint the issue if the system’s performance declines? Since many of these components (e.g., LLMs) are black-box, it is hard, even infeasible, to debug them in isolation. An alternative is to use a \emph{grey-box} approach~\cite{grey-box-testing}: make changes to the system, monitor its performance, and attribute the issues to a particular component. In particular, we can replace a component with an equivalent one (e.g., substitute one model for another, or one vector store for another) or change a component’s configuration (e.g., modify an LLM's temperature or the number of items retrieved by a vector store). We then monitor the performance of the modified system, compare it with the performance of the unmodified system, and use statistical methods to attribute performance regressions to specific components. An exciting research direction is developing statistical techniques that can accurately identify problematic components with minimal additional data and suggest potential changes to improve performance.

{Appendix C} provides several suggestions of projects that focus on verifiability and debuggability.

\subsection{Modularity}
\label{sec:modularity}

To implement more sophisticated tasks, practitioners and researchers have developed frameworks that combine simpler components into larger, more complex systems. Examples of such frameworks include LangChain~\cite{LangChain}, LlamaIndex~\cite{LlamaIndex}, AutoGPT~\cite{AutoGPT}, FrugalML~\cite{FrugalML}, FrugalGPT~\cite{FrugalGPT}, MemGPT~\cite{MemGPT}, and others. Additionally, a recent blog post on compound AI systems demonstrates how composing LLM-based components can significantly enhance performance across various tasks~\cite{compund-AI}. All these efforts are closely related to the concept of modularity. Modularity involves breaking down a complex task into simpler subtasks, implementing and testing these subtasks independently, and then combining the components that implement them into a cohesive system.

Clear statement specifications are essential for modularity. Without each component having a clear statement specification, it becomes difficult, even infeasible, to define the behavior and outputs of the entire system. Therefore, the approaches to improving statement specifications, which we discussed in Section~\ref{sec:sol-prompt-disambiguation}, will directly contribute to enhancing modularity.

Next, we propose two additional approaches for building more effective modular LLM-based systems.

\textbf{Task decomposition:} Given a complex task, how should we decompose it into smaller tasks to optimize performance, cost, or a combination of both? Early work in this area includes {FrugalML}~\cite{FrugalML} and {FrugalGPT}~\cite{FrugalGPT}, which optimize cost and performance by selecting among similar AI services; {RouteLLM}~\cite{RouteLLM}, which chooses across multiple LLMs based on prompt difficulty; {Mixture of Agents}~\cite{MixtureOfExperts} and {Network of Networks}~\cite{network-of-networks}, which utilize multiple LLMs, where each LLM can play a different role (e.g., generator, judge, aggregator, majority voter) to boost performance; {AutoGen}~\cite{AutoGen} which decomposes tasks into sub-tasks and implements each sub-task as an agent. These are important lines of work that will help us better understand the behavior of compound AI and agent-based systems. This understanding can be used to devise new algorithms and techniques for architecting complex systems by defining the key components and how they connect to one another, similar to how we design and architect software systems today.

Furthermore, instead of the developer architecting the system, we could have another LLM-based solution do this. Such a solution could experiment with different decompositions and components to find the optimal configuration.

\textbf{Increasing diversity:} Task specification and verification are not enough—we also need to find an actual solution to the task. One approach to finding a solution is to generate a large number of proposals and check each one until we find one that passes the task's solution specification. To increase the chances of finding such a solution, we can maximize the diversity of the generated proposals. Many problem-solving LLM-based systems, such as {AlphaCode}~\cite{AlphaCode} and {AlphaGeometry}~\cite{AlphaGeometry}, (implicitly) rely on this property. Diversity is also a key of widespread approaches such as Monte-Carlo Tree Search and Reinforcement Learning (e.g., {AlphaGo}~\cite{AlphaGo}, {Alpha Go Zero}~\cite{AlphaGoZero}) which rely heavily on exploration to find new, better solutions. These approaches achieve diversity by carefully exploring the solution space. One research direction is to efficiently generate diverse proposals for general tasks.

\subsection{Reusability}
\label{sec:resuability}

Reusability refers to the ability to leverage already \emph{existing} components to build larger, more complex systems. It is closely tied to modularity, which specifies that a system is composed of multiple components. Reusability complements this concept by incorporating pre-existing components into modular systems. This is an area where LLMs excel. Foundation models, such as ChatGPT and Claude, are frequently offered as hosted services, and the majority of current LLM applications rely heavily on these services. However, there have been relatively few successes in combining these services to create larger, more powerful systems. We argue that this is primarily due to the lack of clear specifications, which complicates reasoning and debugging in such systems. This observation highlights that disambiguating statement specifications could significantly enhance the reusability of existing components.

Next, we explore two additional directions for future work.

\textbf{Constraint-based decomposition:} When decomposing a task into smaller, simpler components, we often want to reuse existing components that we have built or have access to, such as a specific LLM model or a particular agent. One question that arises is which of the existing components should be used, and how to best integrate these components into the overall system architecture.

\textbf{Component substitution:} Assume we have already built an LLM system and now want to substitute one component with another—perhaps we’ve changed the LLM provider, or a new, improved version of a model we are using has been released. Can we replace the LLM used by an application with another LLM without negatively impacting the application’s performance? More generally, can we predict the performance of the application after replacing the LLM?

\subsection{Automated decision making}
\label{sec:adm}

In contrast to traditional software, nearly all of today’s generative AI applications rely on a human in the loop—-whether in text-to-image and text-to-video generation, customer support, programming assistants, summarization, document search, or other domains. As discussed, this is because LLM prompts generally lack clear and precise solution specifications, making it challenging to assess the correctness of their outputs.
However, even when tasks have unambiguous specifications, humans may still be required to assess their correctness if they have context that is missing from specifications (see discussion in Section~\ref{sec:verifiability}).

Thus, to enable automated decision-making, we need not only clear solution specifications but also the ability to verify these specifications automatically. This requires codifying any human context needed for verifiability within the solution specification. Likewise, any tools required for verification—--such as calculators or virtual machines to execute generated code—--must be invoked automatically, with their results interpreted unambiguously.

Possible research directions include: (1) assessing when a solution specification can be verified automatically, (2) identifying gaps that hinder automatic verification of a solution specification, and (3) developing methods to address and fill these gaps.

\section{Summary}
\label{sec:summary}

Engineering disciplines, including software engineering, have been responsible for driving staggering economic progress since the start of the industrial revolution. In this paper, we contend that this progress was enabled by five key properties: verifiability, debuggability, modularity, reusability and automatic decision making. The first two properties enable developers to quickly build systems from scratch, the next two properties enable developers to build \emph{reliable} LLM-based systems, and the last property enables developers to build systems that require no human intervention. Altogether, these properties have enabled an exponential growth of the ecosystem which ultimately fueled economic growth.

More importantly, we assert that the foundation for these engineering properties lies in having clear \emph{specifications} that precisely describe what a task should accomplish (i.e., statement specifications) and provide all the information needed to verify whether the task's outputs are correct (i.e., solution specifications).

AI, and in particular LLMs, places us on the cusp of another economic and social revolution. However, we observe that the specifications---both statement and solution specifications---of most LLM tasks are ambiguous, in part because they use natural language. While natural language makes LLMs widely accessible and general (i.e., \textit{anyone} can describe \textit{any} task), its inherent ambiguity makes it difficult to build new, more powerful and reliable systems from (existing) simpler ones, which ultimately hampers the ecosystem and economic growth.

To address this limitation, we argue for developing new techniques that make it easier to write clear statement specifications and solution specifications to verify that an LLM system's outputs are correct, and debug the LLM system if they are not. By doing so, we can support all five software engineering properties, and enable developers to accelerate the development of new LLM solutions, solutions that are more powerful and reliable, and which can address a growing number of use cases.

\paragraph{Acknowledgements:} We would like to thank everyone who reviewed earlier versions of this paper and provided valuable feedback, which significantly improved its quality: Lakshya A. Agrawal, Ali Ghodsi, Zhuohan Li, Robert Nishihara, Ying Sheng and Lianmin Zheng.

\bibliographystyle{ieeetr}
\bibliography{refs}
\newpage
\appendix

\section*{Appendix A: The five engineering properties: Software Systems vs LLMs}
\label{sec:appendix-A}

Table~\ref{tbl:five-properties} compares how software systems and LLM systems satisfy the five properties of software engineering (see \cref{sec:five-properties}). 

\begin{table}[h!]
\centering
\begin{tabular}{|p{2cm}|p{2.6cm}|p{4.7cm}|p{4.5cm}|}
\hline
\textbf{Property} & \textbf{Description} & \textbf{Software Systems (mostly unambiguous, but less accessible (need to learn a programming language) and less general (only support computation tasks))} & \textbf{LLM Systems (mostly ambiguous, but more accessible (just describe the task in english), and more general (support any task one can describe in english))} \\
\hline
\textit{Verifiability} & If output meets the specification, the output is valid; otherwise it is invalid & Specifications enable output validation, e.g., via validation tests, proof that the solution satisfies specification.  & If a task is ambiguous (e.g., natural language), it is very hard to validate the output given an input. Typically humans verify the output is valid by using extra context or/and additional tools. \\
\hline
\textit{Debuggability} & If output is not correct, modify the system to generate the correct output (i.e., output to meet program’s specification) & Programs are recursively built from smaller components each with a unambiguous specification. This enables debugging by recursively narrowing down the segment of code responsible for the incorrect output and fixing the code. & LLMs are black-boxes, which makes it very hard to debug: you can only change the prompt and there is no easy way to reason on how the input can change the output.  \\
\hline
\textit{Modularity} & Build bigger systems from smaller ones.  & Speed up building larger apps from smaller components which are implemented and tested in parallel (e.g., microservices, workflows, multi-module program). & Enable developers building more complex apps by composing multiple LLMs. Unfortunately, if these LLMs are ambiguous, the outputs of the composed app will also be ambiguous. \\
\hline
\textit{Reusability} & Reuse existing components to build new ones. & Speed up building new apps by leveraging existing components (e.g., libraries, databases, web servers, load balancers). & Enable building new apps leveraging existing components (e.g., Vector Stores) and LLM service (e.g, ChatGPT).   \\
\hline
\textit{Automated decision making} & Build systems that make decisions without human intervention.  & Enable building automated decision-making systems  by leveraging unambiguous and complete  specifications to guarantee the correctness of the output, e.g., car, plane, factory software.  & Most popular LLM apps have a human in the loop. This is because, in general, specifications are ambiguous so cannot guarantee output correctness. \\
\hline
\end{tabular}
\caption{Comparison between traditional software development and LLM prompting.}
\label{tbl:five-properties}
\end{table}

\section*{Appendix B: Why modular AI systems?}
\label{sec:appendix-B}

\begin{figure}[h]
    \centering
    \includegraphics[width = \textwidth]{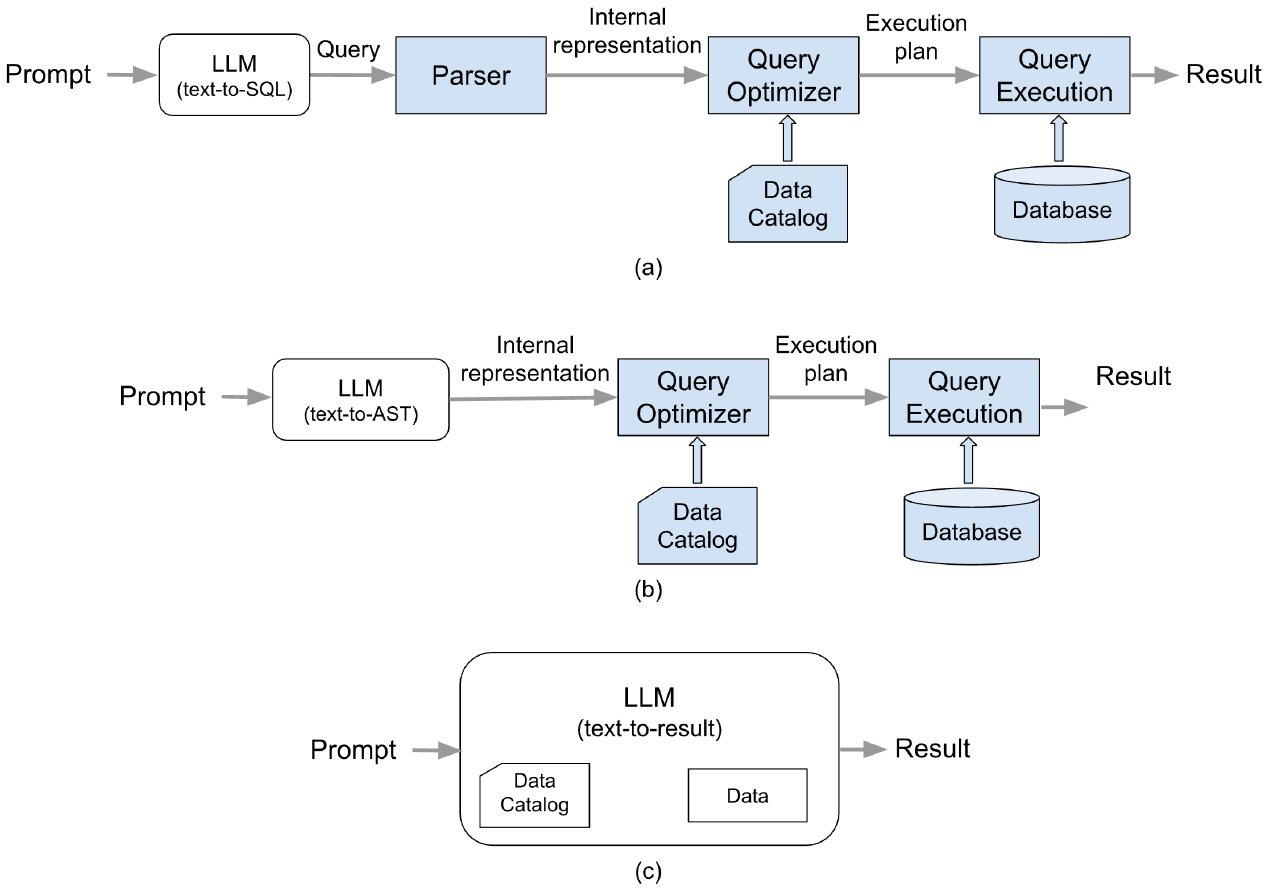}
    \caption{Different options to implement a query optimizer that takes as input the query description in natural language: (a) translate text to SQL and then use an existing query engine; (b) translate text to intermediate representation and then use existing query optimizer and execution components; (c) implement everything in an LLM (here we assume the LLM was trained on date and metadata or these are part of its context/prompt)}
    \label{fig:query-llm} 
\end{figure}

So far we have implicitly assumed that it is a good idea to build AI systems by combining many sub-systems (components). But why is this? After all, the generality of LLMs enables anyone to specify any task. So, why not just describe that task carefully in a prompt and use a single LLM call to solve it? Here are few reasons (many of them have been well articulated in a recent blog on {Compound AI Systems}~\cite{compund-AI}) of why we might want to build modular AI systems:

\begin{itemize}
\item \textbf{Better accuracy}: Given a particular domain, it is possible to use specific data to train or fine-tune a model to optimize its performance for that domain. Furthermore, recent research has shown that it is possible to provide better accuracy by aggregating multiple calls to different LLMs or the same LLM (see \href{https://arxiv.org/pdf/2407.16831}{here}). In particular, recent research has shown that one can get more reliable results by decomposing a complex task into smaller tasks, solve each of these tasks independently, and aggregate the results.   
\item \textbf{Lower cost}: By fine-tuning models on domain specific data it is possible to achieve the same performance or even exceed the performance of a generic frontier model such as ChatGPT at a fraction of cost. This also lowers the development cost by allowing one to stitch together existing LLMs and create more powerful systems.  
\item \textbf{Faster improvements}: Decomposing a complex task into subtasks, and having a component implementing each subtask, makes it easier to debug and improve the results. This is because it is easier to (1) identify which component doesn’t perform well, (2) fix a component that implements a specialized rather than a general task, and (3) reuse existing components that already implement the desired subtask.  
\item \textbf{Integration with software systems}: A modular system can much easier to mix LLM components with existing software components using the best component for the subtask at hand. For example, for arithmetic operations a simple calculator can be much faster and far more accurate than the most powerful LLM.  Figure~\ref{fig:query-llm} shows several possibilities of extending our query engine from Figure~\ref{fig:query-optimizer} to take the query description in natural language.
\end{itemize}

\section*{Appendix C: Research projects}
\label{sec:appendix-C}

\begin{table}[h!]
\centering
\begin{tabular}{|p{2cm}|p{4.4cm}|p{4.2cm}|p{3.2cm}|}
\hline
\textbf{Project} \newline & \textbf{Description} & \textbf{Properties} & \textbf{Comments} \\
\hline
\textit{Iterative disambiguation} & Iteratively check for inputs that generate different outputs and add rules to disambiguate these outputs, as needed. & Disambiguate specification &  \\
\hline
\textit{Domain-specific rules} & Explicitly extract or define the rules for the domain and use these rules to disambiguate the specification and verify the implementation. & Verifiability: use rules to check the implementation & A rule is akin to a unit test. \\
\hline
\textit{formal specification → code + proofs} & E.g., (1) the user uses Coq/Gallina (or similar) to write formal specifications; (2) the LLM generates both programs and proofs that the program meets the certification. & Verifiability: verify proof \& Debuggability: If the proof verification fails, the way it fails might help the users identify bugs &  \\
\hline
\textit{Tests → code} & Given a comprehensive set of tests, generate the code to pass these tests. & Verifiability: via testing \& Debuggability: If a test fails, it might help the user identify bugs & Similar to DSPy for code generation, and generate code from input/output examples in software systems \\
\hline
\textit{Fact-based validation} & Output provides references to reputable sources  & Verifiability: verify the information in the output is correct. Debuggability: if the output answers the prompt but the references are inconsistent with the output, we can identify the problems with the output.  & Verifying the information in the output is correct, doesn’t necessarily mean the answer is correct (the output might have no relation with the input). \\
\hline
\textit{Consistency checking (generate both code and tests)} & Generate the solution (program) and tests from the same prompt. If solution passes the tests, we assume the solution is correct,  & Verifiability: Improves results, but no guarantees. Debuggability: if tests are correct. In this case, the failed tests can point us to bugs in the code. & Code might pass tests, but both code and tests can be wrong. \\
\hline
\textit{(Automatic) prompt engineering} & Chain-of-X, self-notes, memGPT, DSPy (automatic), etc  & Debuggability: Improve results.  & Orthogonal to other projects. \\
\hline
\textit{Hybrid Systems} & Compare Monolithic LLM systems with systems that use modular LLM components, and Hybrid systems that combine LLM modules with “Good Old Fashioned Engineering (GOFE)” modules such as search engines (RAG), FEM simulations, FFTs, linear and non-linear regression, etc. & Verifiability: GOFE components can be used in parallel to verify outputs. Debuggability: GOFE modules from libraries have been extensively debugged.  &  \\
\hline
\end{tabular}
\caption{Several projects to help LLMs provide better support for the five properties in Table~\ref{tbl:five-properties}.}
\label{tbl:llm-projects}
\end{table}

Table~\ref{tbl:llm-projects} shows several research projects focussed on verifiability and debuggability.

\section*{Appendix D: Frequently asked questions}
\label{sec:appendix-D}

\textbf{LLMs are still making rapid progress so is modularity necessary?} Indeed, for the past decade every year or so we have witnessed the emergence of new models that significantly outperforms and replaces the previous models. For example, consider the sequence of revolutionary models released by OpenAI over the past six years: GPT 1 in 2018, GPT 2 in 2019, GPT 3 in 2020, GPT 3.5 in 2022, and GPT 4 in 2023. So wouldn’t this process be enough to lead to exponential growth? While this might be the case, it does not change the fact that without modularity and reusability LLMs will be deprived of a huge vector of growth which is enabled by modularity/reusability. 

Car industry might provide a good analogy. There have been huge advancements from the introduction of the {first car}~\cite{first-car} in 1886 until the introduction of {Ford Model T}~\cite{ford-t} in 1908. During this time period new cars were manually built from the ground up similarly with today's LLMs, and each generation was markedly better then the previous one. But only after the introduction of the assembly line (which basically leverages modularity) to produce Ford Model T did the car industry explode.

\textbf{Aren’t humans effective in collaborating despite their ambiguous interactions?} Humans are using natural language to communicate, which is inherently ambiguous. Despite this, they are still able to work effectively in organizations, so isn’t this a form of modularity? Why wouldn’t we be able to successfully compose LLMs in a similar way? While this is true to some extent, scaling human organizations has been an exercise in removing the ambiguity rather than tolerating it. Indeed, as a company scales, one needs to add more structure in the form of clearly defining different organizations (i.e., product, engineering, marketing, sales, finance, etc) and processes (e.g., financial reports, product planning, tracking key initiatives using objectives and key performance indicators, etc). All this added structure has one role: remove the ambiguity. In fact, the most scalable human organization is arguably the army. But the army is also the organization that defines the most unambiguous communication standards and how everyone should behave and act given her position in the command chain. Thus, the history shows that reducing ambiguity is key to scaling human organizations, which is the main point of this paper. 

\textbf{Aren’t the ambiguous specifications just a reflection of the task people are using the LLMs for rather than the LLMs themselves?} We believe the ambiguity is directly related to the difficulty of specifying a task, rather than an inherent property of the task, i.e., with enough effort most tasks can be significantly disambiguated. In the end, for any given prompt, the user still needs to make a decision of whether the response is good or not. The problem is that the user makes the decision on internal context not available to the LLM. However, this does not mean this context cannot be externalized. To see this, consider that you want to generate an image based on a text prompt, e.g., “Golden Gate Bridge at sunrise”.  Now, imagine that you cannot choose the picture yourself; instead the picture is picked by someone else. What would you do then? Well, probably you are going to give that person as much information as you can to increase the probability of that person will pick a picture you will like, e.g., the picture should use pastel colors, the sun should raise from the city, there should be realistic proportions between the bridge and buildings, etc. All this information helps to disambiguate the task. 

\textbf{What are some examples of specifications in engineering?} In general, there are two types of specifications. The first type targets system’s builders. These specifications should be detailed enough so one can actually build a system using its specification (see Figure~\ref{fig:query-llm}(a)). The second type targets the system's users. These specifications should be detailed enough so one can operate the system. For physical systems such as cars, TVs, phones, microprocessors, storage systems, such specifications often come in the form of user manuals. These manuals contain instructions like setting up the system, powering it on, switching it off, a description of what each button (if any) does, and what each display (if any) should show (see Figure~\ref{fig:query-llm}(b)). These manuals also describe the expected behavior of the system, possible errors, and troubleshooting steps. Irrespective of the type, a specification defines the outputs a system should produce given the input, which is necessary to achieve the five key properties of software systems.

\begin{figure}[h]
    \centering
    \includegraphics[width = \textwidth]{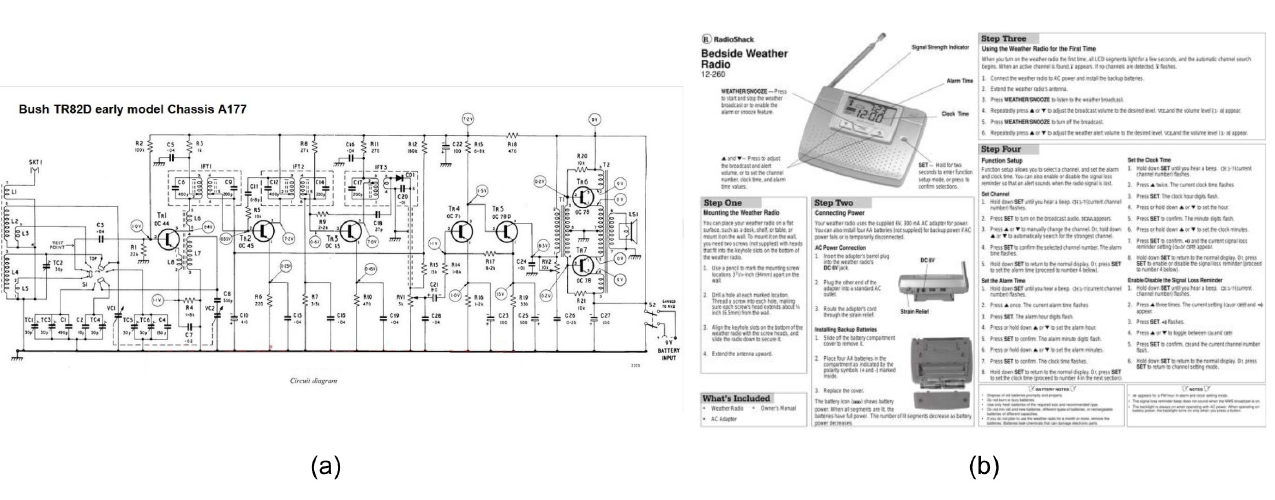}
    \caption{Two radio specifications, (a) for builders and (b) users, respectively}
    \label{fig:spec-examples} 
\end{figure}

\clearpage

\end{document}